\documentclass[12pt]{iopart}

\usepackage{tikz-cd, abraces, iopams, setstack, graphicx, mathrsfs, xcolor, soul}

\newcommand{\up}{\uparrow}
\newcommand{\down}{\downarrow}
\newcommand{\eqref}{\eref}

\begin{document}

\title{Fluctuation relations for a few observable currents at their own beat}

\author{Alberto Garilli, Pedro E. Harunari and Matteo Polettini}

\address{Complex Systems and Statistical Mechanics, Department of Physics and Materials Science,
University of Luxembourg, L-1511 Luxembourg, Luxembourg} 

\ead{alb.garilli@gmail.com}

\date{\today}

\begin{abstract}
Coarse-grained models are widely used to explain the effective behavior of partially observable physical systems with hidden degrees of freedom. Reduction procedures in state space typically disrupt Markovianity and a fluctuation relation cannot be formulated. A recently developed framework of transition-based coarse-graining gave rise to a fluctuation relation for a single current, while all others are hidden. Here, we extend the treatment to an arbitrary number of observable currents. Crucial for the derivation are the concepts of mixed currents and their conjugated effective affinities, that can be inferred from the time series of observable transitions. We also discuss the connection to generating functions, transient behavior, and how our result recovers the fluctuation relation for a complete set of currents.
\end{abstract} 


\section{Introduction}

\def\shift{45pt}
\def\figlabeline{(0,25pt)}

\begin{figure}[ht!]
	\centering
	\includegraphics[width = \textwidth]{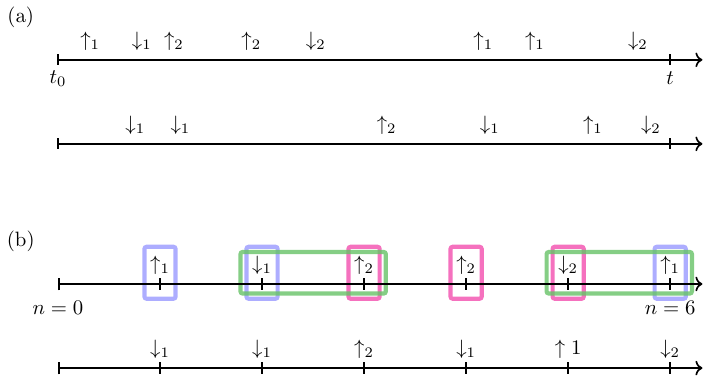}
	\caption{(a) Examples of sequences of observable transitions in the two possible orientations $\up$ or $\down$ obtained from a system with two observable transitions up to a time $t$. Each outcome can have a different number of observed transitions. (b) The same process can observed up to the occurrence of $n=6$ observable transitions. The observables we are interested at are extracted from the so obtained sequences by counting the occurrences of transitions of each kind (blue for kind 1 and pink for kind 2) and the mixed occurrences, in which a transition of one kind follows a transition of a different kind (highlighted in green).}
	\label{fig:trajectoryvisual}
\end{figure}

Consider an experiment on a system whose internal dynamics is understood to be Markovian on a finite set of states. Such a simple process can be thought of as living in a graph with each node representing a state of the system and the edges representing kinds of forward and backward transitions between pairs of states. Here, a kind refers to one edge of the graph irrespective of its orientation, and we denote a forward transition of kind $\nu$ by $\uparrow_\nu$ while $\downarrow_\nu$ is the backward. Suppose that we measure extensive physical quantities associated with certain transitions between different pairs of states, up to the elapse of a fixed clock time $t$. The output of such an experiment will be a time series of observable transitions, see e.g. \fref{fig:trajectoryvisual}(a). The main objective of this contribution is the generalization of the single observable current fluctuation relation, proven by the Authors in Ref.\,\cite{beatofacurrent}, to an arbitrary number of observable currents. For a ``complete'' set of currents the result reproduces a well-known fluctuation relation \cite{Andrieux_2007}.

From a Schnakenberg cycle analysis perspective \cite{schnakenberg,avanzini2023methods}, a complete set of currents covers all cycles of a graph in the sense that the removal of the edges supporting the currents results in a spanning tree with no cycles. Otherwise, if some cycles survive, we say that the set of currents is partial. It is also said to be complete because stationary currents flowing through any edge of the graph can be obtained by a linear combination of elements in the complete set. In this case, the fluctuation relation (FR) for currents
\begin{equation}
\ln \frac{p_t (\lbrace c_\nu \rbrace)}{p_t (\lbrace - c_\nu \rbrace)} = \sum_\nu a_\nu c_\nu 
\label{eq:FRandrieux}
\end{equation}
holds, where $p_t(\bullet)$ denotes the probability of quantities obtained from a $t$-long trajectory, $a_\nu$ are cycle affinities, $c_\nu$ are the currents, and the sum runs over such a complete set. Equality in the equation above is reached under the choice of a preferential initial probability in state space \cite{transient}. The quantity on the right-hand side (RHS) is sometimes called the entropy flow, differing from the entropy production by a boundary term that accounts for initial and final states. 

A partial set of observable currents generally does not satisfy the relation above. To describe the dynamics of a few transitions, one typically relies on coarse-grained models, which generally entail the loss of Markovianity \cite{Esposito_2012,Bo_2017}, which is cured with assumptions on the relaxation timescales, restricting the range of its applicability. Moreover, a thermodynamically consistent description cannot be constructed, since the validity of FRs provides fluctuation-dissipation relations when close to equilibrium \cite{Andrieux} and generalizes the second law. FRs for partial sets of observables evaluated at clock time $t$ have been studied and obtained in several works Refs.\,
\cite{hartich2014stochastic, shiraishi2015fluctuation, rosinberg2016continuous,  crooks2019marginal, marginal, effectivefluc}. However, they require the definition of auxiliary dynamics, whose operational realization is not always granted.

Our strategy in this article is to lift the description from state space to transition space by employing a coarse-graining scheme based on the occurrence of observable transitions \cite{harunari, vandermeer_2022}. We explore details of processes in transition space and recover a FR for partial sets, with the main ingredient being the observation of currents up to a fixed number $n$ of occurrences, namely at their own beat (see \fref{fig:trajectoryvisual}(b)). Unlike the case of a single observable current \cite{beatofacurrent}, we show that it is also necessary to keep track of quantities, coined mixed currents, accounting for the sequences of transitions over distinct edges (see the green boxes in \fref{fig:trajectoryvisual}(b)). The general philosophy is thus that there is a give-and-take between space and time: one can renounce to ``completeness'' of information by including some more memory, which is encoded by observables that account for the previous occurring transition, despite the dynamics being Markovian.

\subsection{Plan of the paper}

This paper is organized as follows: In \sref{sec:2transitions} we first imagine an experimenter monitoring two edges (two kinds of transition) to check for a FR depending on whether they form a complete set or not. In \sref{sec:statement}, we briefly state the results of this paper. The rigorous treatment starts with \sref{sec:setup}, where we introduce the notation and the main definitions which allow in \sref{sec:results} the generalization of the theory to an arbitrary number of observable transitions. The derivation of the FR for currents and mixed currents is in \sref{sec:FR}, while the FR for complete sets of currents is in \sref{sec:completecurr}. In \sref{sec:pref}, we derive the preferred initial distributions that provide transient FRs. We discuss the results and conclude in \sref{sec:conclusions}.

\section{An experiment with two observable currents}

\label{sec:2transitions}

Before introducing the general theory for an arbitrary set of observable currents, we show an example where two currents are observable, and make some initial considerations on the information that is possible to extract by observing time-ordered sequences of observable transitions.

\begin{figure}
	\centering
	\includegraphics[width=.7\textwidth]{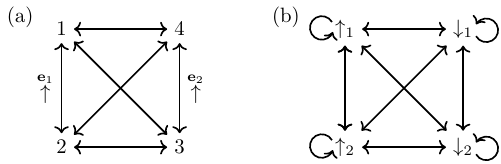}
	\caption{(a) A graph with 4 states and 6 edges, where 3 currents are necessary to form a complete set. Edges denoted with $\mathbf{e}_1$ and $\mathbf{e}_2$ are accessible to observation. The arrows denote the conventional positive sign for the currents, which is denoted, for each edge, as $\up_\nu$, with $\nu = 1,2$. (b) The corresponding graph in the space of observable transitions is completely connected to ensure that each trajectory in this space can be time-reversed. For the same reason, the loops connecting observable transitions in one direction to themselves appear.}
	\label{fig:model}
\end{figure}

Let us suppose that the underlying system dynamics is described by a Markov process on the graph in \fref{fig:model}(a). The nodes represent different states of the systems, and the edges represent the possible transitions between pairs of states. We assume that only the transitions between the pairs of states $(1,2)$ and $(3,4)$ can be independently tracked, and that they are observable in both directions. Referring to \fref{fig:model}(a), we indicate with $\up_1$ transition $2\to 1$ and with $\up_2$ transition $3\to 4$, while $\down_1$ and $\down_2$ indicate the reverse transitions $1\to 2$ and $4 \to 3$ respectively. We indicate an observable transition with symbol $\ell_\nu$, where $\nu = 1,2$ labels its kind (i.e. the edge where the transition occurs). The experimenter then collects sequences of transitions $\mathcal{L}_n : =\{\ell^{(1)}, \ell^{(2)}, \ldots \ell^{(n)}\}$ by stopping the sampling after a given number $n$ of observed transitions, obtaining outcomes such as the ones illustrated in \fref{fig:trajectoryvisual}(b).

Let us now indicate by $\ell$ (without index) a generic observable transition. The probability $p(\ell|\ell')$ to observe $\ell$ after another transition $\ell'$ has occurred is called the trans-transition probability \cite{beatofacurrent}, and can be derived given knowledge of the transition rates of the full system \cite{beatofacurrent,harunari}. Such probabilities can be arranged in a trans-transition matrix
\begin{equation}
P = \left(\begin{array}{c c c c}
p(\up_1|\up_1) & p(\up_1|\down_1) & p(\up_1|\up_2) & p(\up_1|\down_2)\\
p(\down_1|\up_1) & p(\down_1|\down_1) & p(\down_1|\up_2) & p(\down_1|\down_2)\\
p(\up_2|\up_1) & p(\up_2|\down_1) & p(\up_2|\up_2) & p(\up_2|\down_2)\\
p(\down_2|\up_1) & p(\down_2|\down_1) & p(\down_2|\up_2) & p(\down_2|\down_2)
\end{array}\right)
\label{eq:trans-transition2kinds}
\end{equation}
which describes a discrete-time Markovian process in the space of observable transitions \cite{beatofacurrent} represented by the graph in \fref{fig:model}(b), with $p_k(\ell) = \sum_{\ell'} P_{\ell,\ell'} p_{k-1}(\ell')$, where $p_k(\ell)$ denotes the probability of the $k$-th observed transition being $\ell$, for $k = 1, \ldots, n$.

Given a sequence $\mathcal{L}_n$, the experimenter can extract the number of times $n_{\ell_\nu}$ a transition $\ell_\nu$ occurs ($\nu$ representing its kind) and the number of times $n_{\ell_\nu\ell_\mu'}$ a transition $\ell_\nu$ occurs after $\ell_\mu'$. The following current-like observables can be obtained from these countings:
\begin{enumerate}
\item The total currents $c_\nu(\mathcal{L}_n) = n_{\up_\nu}(\mathcal{L}_n) - n_{\down_\nu}(\mathcal{L}_n)$ of same kind $\nu = \{1,2\}$;
\item The loop currents $\xi_\nu(\mathcal{L}_n) = n_{\up_\nu\up_\nu} (\mathcal{L}_n)- n_{\down_\nu\down_\nu}(\mathcal{L}_n)$ of same kind $\nu = \{1,2\}$;
\item The mixed currents $\xi_{\ell_1\ell_2}(\mathcal{L}_n) = n_{\ell_1\ell_2}(\mathcal{L}_n) - n_{\bar{\ell}_2\bar{\ell}_1}(\mathcal{L}_n)$, with $\ell_1 \in \{\up_1,\down_1\}$ and $\ell_2 \in \{\up_2,\down_2\}$, which by convention flow from kind 2 to kind 1.
\end{enumerate}
Above, we have used the symbol $\bar{\ell}$ to identify the reversed transition. Also notice that loop and mixed currents are evaluated by taking the differences of each $n_{\ell\ell'}$ with $n_{\bar{\ell'}\bar{\ell}}$, as we adopt a notion of time-reversal where both the order and the orientation of transitions in $\mathcal{L}_n$ are inverted.

Having access to all this information, the experimenter is interested in checking whether the joint statistics of the observable total currents $c_1,c_2$ satisfies a FR at large $n$, and if not, what additional information is necessary to reinstate them. In fact, it was recently proven by the Authors that an observable current $c$ associated with a single observable transition $\ell\in\{\up,\down\}$ satisfies the asymptotic FR when evaluated after $n$ occurences \cite{beatofacurrent}
\begin{equation}
    \ln\frac{p_n(c)}{p_n(-c)} = a c ,
\label{FRsingletrans}
\end{equation}
with 
\begin{equation}
    a = \ln \frac{p(\up|\up)}{p(\down|\down)}
\end{equation}
the so-called effective affinity \cite{marginal,neri}. The relation above is valid at all finite $n$ by a specific choice of probability distribution over the initial transition, otherwise it shows an additional term accounting for the transitions at the boundaries.

However, the joint statistics of $c_1,c_2$ alone turns out to generally not satisfy a FR. In fact, the experimenter is not using all the information that can be extracted from sequences $\mathcal{L}_n$, which also includes loop currents and mixed currents. 

In this work, therefore, we extend  Eq.\,\eqref{FRsingletrans} to the case of multiple observable currents by showing that a FR is satisfied by complementing the statistics of observable currents with the mixed currents, which can be inferred from sequences of transitions. For two observable currents $(c_1,c_2)$ we find that
\begin{equation}
\fl \ln \frac{p_n(c_1,c_2,\boldsymbol{\xi}, \ell^{(1)},\ell^{(n)} )}{p_n( - c_1, - c_2, - \boldsymbol{\xi}, \bar{\ell}^{(1)},\bar{\ell}^{(n)})}
= a_1 c_1 + a_2 c_2 + \boldsymbol{\tilde{\alpha}} \cdot \boldsymbol{\xi} + \Delta \tilde{u}(\ell^{(1)},\ell^{(n)})
\label{eq:FRexperiment}
\end{equation}
with effective affinities $a_1,a_2$ respectively conjugated to the total currents $c_1$ and $c_2$, and mixed affinities $\boldsymbol{\tilde{\alpha}} = \{ \tilde{\alpha}_{\up_1\up_2}, \tilde{\alpha}_{\down_1\up_2} , \tilde{\alpha}_{\up_1\down_2} , \tilde{\alpha}_{\down_1\down_2} \} $ (see \sref{sec:FR} for details) conjugated to the mixed currents $\boldsymbol{\xi} = \{ \xi_{\up_1\up_2},\xi_{\down_1\up_2},\xi_{\up_1\down_2},\xi_{\down_1\down_2} \}$. Moreover, the expression above shows a bounded additional term $\Delta \tilde{u}(\ell^{(1)},\ell^{(n)}) := \tilde{u}(\bar{\ell}^{(n)}) - \tilde{u}(\ell^{(1)})$ that accounts for the initial and final transitions in a sequence $\mathcal{L}_n$. The affinities are defined from the trans-transition probabilities contained in the trans-transition matrix Eq.\,\eqref{eq:trans-transition2kinds}, and therefore, if an ergodic principle holds (see \sref{sec:conclusions} for a discussion), they can be estimated off-shell from another experiment where a long sequence of observable transitions is collected. When considering complete sets of currents the term $\boldsymbol{\tilde{\alpha}} \cdot \boldsymbol{\xi}$ can be incorporated into the boundary potential and becomes itself bounded, thus its contribution can be neglected at long times. With a sufficiently long time series, the experimenter empirically estimates all of said quantities using the time series of observable transitions, and can verify that Eq.\,\eqref{eq:FRexperiment} holds.

As a final remark, the number of mixed currents can be reduced by applying Kirchhoff's Current Law (KCL) in the space defined by the observable transitions. This point will be addressed later in \sref{sec:redundancy}.

\section{Statement of the main results}

\label{sec:statement}

Let $\nu$ label different observable undirected edges of the graph where the stochastic dynamics is defined, called the kind. Each observable transition $\ell_\nu$ of kind $\nu$ is allowed in the two possible directions $\ell_\nu \in \{\up_\nu, \down_\nu\}$. Given the ignorance about the details of the system, one may wonder whether the number of observed transitions is sufficient to establish a thermodynamically consistent description of the reduced system. The total current $c_\nu$ along a single edge $\nu$ is obtained by counting how many times the transitions $\up_\nu$ and $\down_\nu$ occur and taking their difference. When the currents are evaluated for a single sequence $\mathcal{L}_n$ at the occurrence of a fixed number $n$ of observable transitions, and when the mixed currents $\xi_{\ell_\nu\ell_\mu} = n_{\ell_\nu\ell_\mu} - n_{\bar{\ell}_\mu\bar{\ell}_\nu},\, \mu>\nu$, are also extracted from $\mathcal{L}_n$, the joint probability distribution $p_n (\lbrace c_\nu \rbrace, \lbrace \xi_{\ell_\nu\ell_\mu} \rbrace)$ satisfies the symmetry
\begin{equation}
\eqalign{
\fl \ln \frac{p_n (\lbrace c_\nu \rbrace,\lbrace \xi_{\ell_\nu\ell_\mu} \rbrace,\ell^{(1)},\ell^{(n)})}{p_n(\lbrace - c_\nu \rbrace,\lbrace -\xi_{\ell_\nu\ell_\mu} \rbrace, \bar{\ell}^{(1)},\bar{\ell}^{(n)})} = \sum_\nu a_\nu c_\nu  + \sum_{\underset{\mu>\nu}{\mu,\nu}} \sum_{\ell_\nu\ell_\mu} \tilde{\alpha}_{\ell_\nu\ell_\mu} \xi_{\ell_\nu\ell_\mu} + \Delta\tilde{u}(\bar{\ell}^{(n)},\ell^{(1)}) ,
}
\label{eq:FRgeneralintro}
\end{equation}
which represents the main result of this paper. The affinities 
\begin{equation}
a_\nu = \ln \frac{p(\up_\nu|\up_\nu)}{p(\down_\nu|\down_\nu)} ,
\end{equation}
each conjugated with the total current $c_\nu$ are called effective affinities. Affinities $\tilde{\alpha}_{\ell_\nu\ell_\mu}$ are conjugated to mixed currents $\xi_{\ell_\nu\ell_\mu}$ and are defined as
\begin{equation}
\tilde{\alpha}_{\ell_\nu\ell_\mu} = \ln \frac{p(\ell_\nu|\ell_\mu)}{p(\bar{\ell}_\mu|\bar{\ell}_\nu)} - \frac{1}{2} \left( a_\nu j(\ell_\nu) + a_\mu j(\ell_\mu)\right) ,
\end{equation}
where $j(\up_\nu) = +1$ and $j(\down_\nu) = -1$. The difference of potentials $\Delta \tilde{u}$ is bounded and will be described later in the paper.

A second result in this paper concerns the case of complete sets of currents when evaluated after $n$ observable transitions. In this case, there is no need to track mixed currents, and the effective affinities become cycle affinities. Therefore, the joint probability $p_n(\lbrace c_\nu \rbrace, \ell^{(1)},\ell^{(n)})$, marginalized with respect to the mixed currents, satisfies the symmetry
\begin{equation}
\ln \frac{p_n (\lbrace c_\nu \rbrace, \ell^{(1)},\ell^{(n)})}{p_n (\lbrace - c_\nu \rbrace, \bar{\ell}^{(1)},\bar{\ell}^{(n)})} = \sum_\nu a_\nu c_\nu + \Delta \tilde{\tilde{u}}(\ell^{(1)},\ell^{(n)}),
\label{eq:FRcompleteintro}
\end{equation}
which is analogous to the FR Eq.\,\eqref{eq:FRandrieux} proven in Ref.\,\cite{Andrieux}, except that the external clock time $t$ is replaced by the a fixed number of occurrences $n$ of observable transitions.

The last result of this paper extends the results Eqs.\,\eqref{eq:FRgeneralintro} and \eqref{eq:FRcompleteintro} to all $n$ with vanishing boundary term. The same FRs written in terms of the Moment Generating Functions (MGF) for the joint probabilities $p_n(\lbrace c_\nu \rbrace,\lbrace \xi_{\ell_\nu\ell_\mu} \rbrace)$ and $ p_n(\lbrace c_\nu \rbrace)$ allow to find a preferred initial probability in transition space, $\boldsymbol{p_1^*}$ such that both relations are exact at all $n$. In the case where a non-complete set of currents is observable, the preferred initial probability reads
\begin{equation}
p_{1,{\rm nc}}^*(\ell_\nu) = \frac{\exp \left(\frac{1}{2} a_\nu j(\ell_\nu)\right)}{2 \sum_\mu \cosh\left( \frac{1}{2} a_\mu \right)} ,
\end{equation}
and for complete sets
\begin{equation}
p_{1,{\rm c}}^*(\ell_\nu) = \frac{\exp \left(\frac{1}{2}a_\nu j(\ell_\nu) + \upsilon_\nu\right)}{2 \sum_\mu \exp(\upsilon_\mu) \cosh\left( \frac{1}{2} a_\mu \right)} ,
\end{equation}
with $\upsilon_\nu$ a potential associated with kind $\nu$, which will also be described later.

\section{Setup}

\label{sec:setup}

In this section, we set the notation and introduce the quantities and methods needed to derive our results.

\subsection{Continuous-time Markov chains in state space}
Let us consider an oriented connected graph $\mathscr{G}= (X,E,I)$ consisting of $|X|$ vertices and $N = |E|$ oriented edges $\boldsymbol{e}_\nu$, $\nu = 1, \dots , N$ connecting the vertices $x \in X$ through the incidence relation $I: E \to X^2$. The orientation of the edges can be chosen arbitrarily. A continuous-time Markov process on the graph $\mathscr{G}$ is characterized by assigning the transition rates $r(x|y) \geq 0$ from state $y$ to state $x$ along the corresponding edge $\boldsymbol{e}_\nu$. The stochastic dynamics on the graph $\mathscr{G}$ is generated by a rate matrix $R$ with elements $R_{xy} = r(x|y), \; x\neq y$, $R_{yy} = - \sum_{x\neq y} r(x|y) = - r(y)$ where the diagonal elements are identified as the exit rates $r(y)$ from state $y$, ensuring $\sum_{x} R_{xy} = 0, \, \forall \, y \in X$.

Let $q_t(x)$ be the probability that at time $t$ the system is found in state $x$. Initializing the states with probability $q_0(x)$, $q_t(x)$ evolves according to the master equation
\begin{equation}
\frac{d q_t(x)}{dt} = \sum_{y} R_{x,y}\, q_t(y).
\label{eq:master}
\end{equation}
For the Markov chains considered in this paper we always assume irreducibility, i.e. there exists at least one path connecting any pair of states in $X$. Therefore, the process has a unique stationary distribution satisfying $\sum_{y} R_{x,y} q_{\rm st}(y) = 0$.

The process in state space defined here can be rephrased as a discrete-time process by giving up the information about intertransition times. We define the embedded Markov chain as the process described by the following transition matrix
\begin{equation}
\Pi_{x,y} = \cases{ \pi(x|y) & for $x\neq y$;\\
0 & for $x = y$
}
\label{eq:embeddedchain}
\end{equation}
with
\begin{equation}
\pi(x|y) = \frac{r(x|y)}{r(y)}.
\label{eq:transitionprobabilities}
\end{equation}

\subsection{Full transition space}

Let $\ell_\nu \in \lbrace \up_\nu,\down_\nu \rbrace$ denote the transition $\mathtt{s}(\ell_\nu) \to \mathtt{t}(\ell_\nu)$ from the source state $\mathtt{s}(\ell_\nu)$ to the target state $\mathtt{t}(\ell_\nu)$ along the oriented edge $\boldsymbol{e}_\nu \in E$. If $\ell_\nu$ occurs in the direction parallel to the orientation of $\boldsymbol{e}_\nu$ the transition is denoted as $\up_\nu$, otherwise as $\down_\nu$. The reversed transition along edge $\boldsymbol{e}_\nu$ is denoted $\bar{\ell}_\nu$. Being associated with a single edge, the subscript $\nu$ is referred to, in short, as the ``kind" of transition $\ell_\nu$, while the generic transition, regardless of its kind, is indicated simply by $\ell \in \bigcup_{\nu =1}^N \{\up_\nu,\down_\nu\}$. The space $\bigcup_{\nu =1}^N \{\up_\nu,\down_\nu\}$ is called full transition space, and its elements are all the possible transitions occurring in the full system in both allowed directions. As the process in state space is irreducible, the process described in the space of transitions is also irreducible.

\subsection{Observable transition space}

To account for the scenario of observing a subset of transitions, we consider a subset $\mathcal{O} = \bigcup_{\nu = 1}^M \{\up_{\nu},\down_{\nu}\} $, where, for simplicity, the first $M \leq N$ kinds represent the observable transitions.
Our notion of observability for transitions is related with the exchange of physical quantities, e.g. charge, energy, matter, which can be monitored by an experimental apparatus. Emission of photons, changes of protein configurations, and translocation of molecular motors are examples of observable quantities which can be associated with a certain change of state in the system. The remaining transitions are said to be hidden.

The dynamics in the space of observable transitions can be thought of as living on the graph $\tilde{\mathscr{G}} = (\mathcal{O},\tilde{E}, \tilde{I})$, with $\tilde{E}$ the set of edges connecting visible transitions in $\mathcal{O}$, where bidirectionality is assumed, via the incidence relation $\tilde{I}: \tilde{E} \to \mathcal{O}$. As an example, see the transition space graph for two visible transitions illustrated in \fref{fig:model}(b). From now on, it will be implicit that the symbol $\ell \in \mathcal{O}$ denotes a generic observable transition and $\ell_\nu \in \lbrace \up_\nu,\down_\nu \rbrace$ denotes an observable transition of kind $\nu$.

In order to avoid undetermined effective affinities, as occurs with affinities in the presence of absolute irreversibility (unidirectional edges), we assume the following hidden irreducibility property: The graph $\mathscr{G}' =  (X,E',I')$ with $E' = E\setminus \lbrace \boldsymbol{e}_\nu \rbrace$ and $I':E'\to X$, obtained by removing the set of observable edges $\lbrace \boldsymbol{e}_\nu \rbrace$, is also irreducible \cite{beatofacurrent}. With this property, there exists a hidden path from the tip of any observable transition to another one's source, so every transition sequence is possible. Therefore, the observable transition space is fully connected (cf. \fref{fig:model}(b)). Additionally, it ensures the existence of a unique stationary distribution $\boldsymbol{p}_{\rm st}$ such that $p_{\rm st}(\ell) = \sum_{\ell'}P_{\ell,\ell'} p_{\rm st}(\ell')$.
Notice however that it sets a limit to the number of observable transitions, since it is not possible to have an irreducible $\mathscr{G}'$ if $M$ is larger than the number of cycles in $\mathscr{G}$. As a final comment, all results still hold in the presence of absolute irreversibility over edges in the hidden part of the graph if the conditions above are still satisfied.

\subsection{Observable transitions' dynamics}

\label{sec:obstrans}

The Markovian dynamics in state space induces a stochastic dynamics in the observable transition space. Because each current is associated with a single edge, it can be shown \cite{kris-renewal} that such a process is a Markov renewal process in which intertransition times between each $\ell,\ell' \in \mathcal{O}$ are integrated out of the picture, yielding a discrete-time Markov chain.

Notice that the mapping from a process in state space to a process in the observable transition space is not one-to-one as many different state-space trajectories can induce the same transition space trajectory \cite{beatofacurrent}.

The process in the observable transition space is generated by a stochastic matrix $P$, called the trans-transition matrix, which is obtained directly from the original rate matrix $R$ by solving first-passage time problems. In brief, introducing the taboo matrix $\Theta_{ij} = 1 - \sum_{\ell \in \mathcal{O}} \delta_{i,\mathtt{t}(\ell)}\delta_{j,\mathtt{s}(\ell)}$ and the survival matrix $S$ with entries $S_{ij} = R_{ij}  \Theta_{ij}$, the trans-transition matrix $P$ has entries
\begin{equation}
P_{\ell,\ell'} = p(\ell|\ell') = - r(\ell)\left[ S^{-1}\right]_{\mathtt{s}(\ell),\mathtt{t}(\ell')} ,
\label{eq:transtransition}
\end{equation}
which is the conditional probability of observing $\ell \in \mathcal{O}$ given that the previous observable transition was $\ell' \in \mathcal{O}$. As per the hidden irreducibility assumption, all values of Eq.\,\eqref{eq:transtransition} are positive. The taboo matrix has vanishing elements in the positions corresponding to those pairs of states which are connected by an observable transition and 1 otherwise. It serves as a tool to build the survival matrix, which is obtained from removing all rates associated with observable transitions from off-diagonal elements of the rate matrix $R$ and defines the evolution restricted to hidden transitions $\ell \notin \mathcal{O}$. The inverse of the survival matrix emerges in Eq.\eqref{eq:transtransition} after the intertransition time is integrated from 0 to infinity, thus $P_{\ell,\ell'}$ represents the probability independently of the time span between $\ell'$ and $\ell$. More detailed descriptions of the survival matrix and its relation to trans-transition probabilities can be found in Refs.~\cite{beatofacurrent, harunari}.

Letting $p_k(\ell) := p(\ell^{(k)} = \ell)$, $k = 1, \ldots, n$, be the probability that the $k$-th observable transition $\ell^{(k)}$ is $\ell$, its evolution is determined by
\begin{equation}
p_k(\ell) = \sum_{\ell' \in \mathcal{O}} P_{\ell\ell'} p_{k-1}(\ell') = \sum_{\ell' \in \mathcal{O}} [P^{k-1}]_{\ell\ell'} p_1(\ell')
\label{eq:process}
\end{equation}
with $p_1(\ell')$ the probability that the first observed transition is $\ell' \in \mathcal{O}$. The initial distribution can be expressed in terms of the the probability distribution of the initial state $\{q_x(0), \; x \in X\}$ by
\begin{equation}
p_1(\ell) = -r(\ell) \sum_{x \in X} \left[S^{-1}\right]_{\mathtt{s}(\ell),x}q_x(0).
\label{eq:statestotrans}
\end{equation}

\subsection{Trajectories, time-reversal and path probabilities}

\label{sec:time-reversal}

We consider a sequence of observable transitions of length $n$: $\mathcal{L}_n = \ell^{(1)}\to \cdots \to \ell^{(n)} = \lbrace\ell^{(i)}\rbrace_{i=1}^n$, $\ell^{(i)} \in \mathcal{O}$. Its probability is expressed in terms of trans-transition probabilities as
\begin{equation}
p(\mathcal{L}_n) = p_1(\ell^{(1)})\prod_{k = 1}^{n-1} p(\ell^{(k+1)}|\ell^{(k)}) .
\label{eq:pathFWD}
\end{equation}
The notion of time-reversal in transition space is derived from that in state space. Notice that when we reverse trajectories, transitions occur in the reverse order and with the opposite direction. Therefore, the time-reverse sequence of observable transitions is $\bar{\mathcal{L}}_n = \lbrace\bar{\ell}^{(n-i)} \rbrace_{i=0}^{n-1}$. Its probability is always nonzero due to the biridectionality of observable transitions and the hidden irreducibility, thus
\begin{equation}
p(\bar{\mathcal{L}}_n) = p_1(\bar{\ell}^{(n)}) \prod_{k=1}^{n-1} p(\bar{\ell}^{(k)}|\bar{\ell}^{(k+1)}) .
\label{eq:pathBWD}
\end{equation}

\subsection{Swapping matrix and block-antitransposition}

Let $\mathcal{L}_n$ be the outcome of a process with trans-transition matrix $P$. The probability of a forward sequence is obtained as a product of trans-transition probabilities, each accounting for consecutive transitions in $\mathcal{L}_n$. The same is done for the time-reversed sequence $\bar{\mathcal{L}}_n$, where each trans-transition $p(\ell|\ell')$ is replaced with $p(\bar{\ell'}|\bar{\ell})$ up to boundary terms. We look for an operation that connects each trans-transition probability with its time-reversed analogue. Having access to such an operation will be convenient in \sref{sec:pref} to derive preferred initial probabilities.

Denoting with $M$ the number of observable kinds, we define the matrix $J$ with entries
\begin{equation}
J_{\ell_\nu\ell_\mu'} = \delta_{\ell_\nu,\bar{\ell}_\mu'} \delta_{\nu,\mu},
\label{eq:swappingmatr}
\end{equation}
whose diagonal $2\times 2$ blocks are $M$ copies of the first Pauli matrix, and all the others are zero: 
\begin{equation}
J = \bigoplus_{\nu = 1}^M \left(\begin{array}{cc} 0 & 1 \\ 1 & 0 \end{array}\right).
\end{equation}
We call $J$ the swapping matrix since, when applied to vectors, it swaps the entries labeled as $\ell_\nu$ with their time-reversed $\bar{\ell}_\nu$, and when applied to the left or right of a matrix, it swaps pairs $(\ell_\nu,\bar{\ell}_\nu)$ of rows or columns, respectively.

Given a trans-transition matrix, we can define block-antitransposition through the operation
\begin{equation}
P^\perp = J P^\top J ,
\label{eq:antitranspdef}
\end{equation}
with $P^\top$ denoting the matrix transpose of $P$. The matrix $P^\perp$ is, in fact, obtained from $P$ by swapping each transition $\ell$ with $\bar{\ell}$ and their order, i.e. $P_{\ell_\nu\ell'_\mu}^\perp = P_{\bar{\ell}_\mu' \bar{\ell}_\nu}$, as it can be immediately verified by use of Eq.\,\eqref{eq:swappingmatr}. In simple words, the operation $P^\perp$ maps the trans-transition probabilities $p(\ell|\ell')$ to their time-reversal $p(\bar{\ell}'|\bar{\ell})$. Note that the columns of $P^\perp$ are not normalized, and thus it does not describe a Markov chain.

\subsection{Currents and path probabilities}

\label{sec:currents}
 
Let $\mathcal{L}_n$ be a sequence of length $n$ and $\ell_\nu \in \{\up_\nu,\down_\nu\}$. The main quantity we are interested in is the total current of kind $\nu$ defined as
\begin{eqnarray}
c_\nu(\mathcal{L}_n) & := \sum_{k=1}^n \left[ \delta_{\ell^{(k)} , \up_\nu} - \delta_{\ell^{(k)} , \down_\nu} \right] \\
& = n_{\up_\nu}(\mathcal{L}_n) - n_{\down_\nu}(\mathcal{L}_n)
\label{eq:defcurrents}
\end{eqnarray}
with $n_{\ell_\nu}(\mathcal{L}_n)$ denoting the number of times transition $\ell_\nu$ occurs in $\mathcal{L}_n$. The occurrence of a single transition $\ell_\nu$ thus contributes to the current Eq.\,\eqref{eq:defcurrents} by adding the elementary charge $j(\ell_\nu) := \delta_{\up_\nu,\ell_\nu} - \delta_{\down_\nu,\ell_\nu} $.

\begin{figure}
\centering
\includegraphics[scale=1]{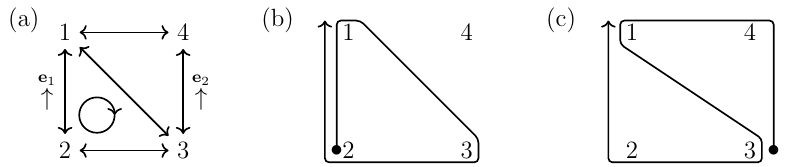}
\caption{(a) A 4 states graph where two currents are necessary to form a complete set. Edges $\boldsymbol{e}_1$ and $\boldsymbol{e}_2$ are observable, thus constituting a complete set. The cycle associated with the chord $\boldsymbol{e}_1$, constructed according to the Schakenberg procedure, is highlighted. (b) For complete sets, the sequence of observable transitions $\up_1 \to \up_1$, where $\up_1$ denotes the transition $2 \to 1$, is associated with the closure of the cycle $1\to 3 \to 2$. (c) The sequence $\up_2 \to \up_1$, with $\up_2: 3\to 4$, is associated with the closure of the same cycle $1 \to 3 \to 2$.}
\label{fig:cyclesloops}
\end{figure}

We aim at deriving a FR for the currents Eq.\,\eqref{eq:defcurrents} at discrete-time $n$. For convenience, we introduce other quantities that can be defined from the conditional numbers $n_{\ell_\nu\ell_\mu'} (\mathcal{L}_n)$, i.e. the number of times transition $\ell_\nu$ occurs after $\ell_\mu'$ in $\mathcal{L}_n$. According to our notion of time-reversal we then define the quantities
\begin{equation}
\xi_{\ell_\nu,\ell_\mu'}(\mathcal{L}_n) = n_{\ell_\nu\ell_\mu'}(\mathcal{L}_n) - n_{\bar{\ell}_\mu'\bar{\ell}_\nu}(\mathcal{L}_n)
\label{eq:xi}
\end{equation}
and
\begin{equation}
\alpha_{\ell_\nu,\ell_\mu'} = \ln \frac{p(\ell_\nu|\ell_\mu')}{p(\bar{\ell}_\mu'|\bar{\ell}_\nu)}.
\label{eq:aff}
\end{equation}
When $\nu = \mu$ and $\ell_\nu = \ell_\mu' = \up_\nu$ the expressions Eqs.\,\eqref{eq:xi} and \eqref{eq:aff} are called the loop currents
\begin{equation}
\xi_\nu(\mathcal{L}_n) = n_{\up_\nu\up_\nu}(\mathcal{L}_n) - n_{\down_\nu\down_\nu}(\mathcal{L}_n)
\label{eq:loopdef}
\end{equation}
and effective affinities conjugated to $\xi_\nu$
\begin{equation}
a_\nu := \alpha_{\up_\nu\up_\nu}= \ln\frac{p(\up_\nu|\up_\nu)}{p(\down_\nu|\down_\nu)} .
\label{eq:affinitydef}
\end{equation}
Referring to Fig.\,\ref{fig:model}(b), loop currents Eq.\,\eqref{eq:loopdef} describe occurrences of subsequent transitions on the same edge with the same orientation, represented by the loops around each vertex in the transition space graph, hence the name. Moreover, in the case of complete sets of currents, this has not to be confused with the amount of times the cycle associated with the chord $\boldsymbol{e}_\nu$ is closed in a single realization of the process (see the example in \fref{fig:cyclesloops} for a visual explanation).

When $\nu \neq \mu$ the quantities defined by Eq.\,\eqref{eq:xi} are called mixed currents $\xi_{\ell_\nu\ell_\mu}$ from type $\mu$ to type $\nu$ and $\alpha_{\ell_\nu\ell_\mu}$ in Eq.\,\eqref{eq:aff} their conjugated mixed affinities. The remaining case where $\nu = \mu$ and $\ell_\nu = \bar{\ell}_\mu$ is of no interest since both expressions vanish. In the following, we arbitrarily choose the sign of the mixed currents such that the positive sign is always associated with passages from kind $\mu$ to kind $\nu$, with $\mu > \nu$.

\subsection{Relation between total currents and loop currents}

Here we derive a linear expression involving the total current $c_\nu(\mathcal{L}_n)$ of kind $\nu$, the loop current $\xi_\nu(\mathcal{L}_n)$ of kind $\nu$ and the mixed currents, in a single realization $\mathcal{L}_n$ of the process in transition space. A similar relation was employed in Ref.\cite{beatofacurrent} to derive the FR for a single observable current. Here, we provide a generalization to multiple observable kinds, which will play a crucial role in the derivation of the FR for multiple observable currents as it allows to define the shifted mixed affinities which, conjugated with the mixed currents, provides the symmetry Eq.\,\eqref{eq:FRgeneralintro} for the joint statistics of the total currents Eq.\,\eqref{eq:defcurrents} and the mixed currents \eqref{eq:xi} for $\mu\neq\nu$.

\begin{figure}
\centering
\includegraphics[width=.9\textwidth]{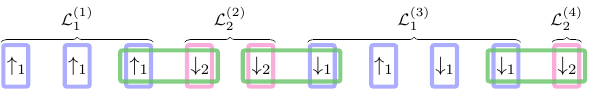}
\caption{The total current in a snippet $\mathcal{L}^{(i)}_\nu$ of type $\nu$ is given by Eq.\,\eqref{eq:totalcurrent}. The total current of type $\nu$ is then the sum over all the snippets $s$ of the same kind, leading to  Eq.\,\eqref{eq:generalizedcurr}.}
\label{fig:snippets}
\end{figure}

Given a sequence $\mathcal{L}_n$ we identify nonoverlapping subsequences $\mathcal{L}_{\nu_i}^{(i)}, \, i = 1,\cdots,m$, in the following referred to as snippets of kind $\nu_i$, containing only consecutive transitions of the same kind. $\mathcal{L}_n$ is then rewritten as a succession of $m$ snippets $\mathcal{L}_{\nu_i}^{(i)}$ of kind $\nu_i$
\begin{equation}
\mathcal{L}_n = \mathcal{L}^{(1)}_{\nu_1} \to \mathcal{L}^{(2)}_{\nu_2} \to \cdots \to \mathcal{L}^{(m)}_{\nu_m}.
\end{equation}
Each snippet $\mathcal{L}_{\nu_i}^{(i)}$ has variable length $n_i$ with $\sum_i n_i = n$, and for convenience, consecutive snippets of the same kind are treated as a single one. \Fref{fig:snippets} represents a sequence of two observable transitions that is cut into snippets of the two kinds as an illustrative example. Each $\mathcal{L}_{\nu_i}^{(i)}$ contributes to the total current of kind $\nu_i$ as \cite{beatofacurrent}
\begin{equation}
c_{\nu_i}(\mathcal{L}_{\nu_i}^{(i)}) = \xi_{\nu}(\mathcal{L}_{\nu_i}^{(i)}) + \frac{1}{2}(j(\ell_{\nu_i}^{(1_i)}) + j(\ell_{\nu_i}^{(n_i)}))
\label{eq:totalcurrent}
\end{equation}
with $\ell^{(1_i)}$ and $\ell^{(n_i)}$ indicating the first and last transition in snippet $\mathcal{L}_{\nu_i}^{(i)}$. Then by summing over all $i'$ such that $\nu_{i'} = \nu$ we obtain the total current of kind $\nu$ in the full sequence $\mathcal{L}_n$ in terms of the loop current $\xi_\nu$ and the mixed currents as (see \sref{app:1} for a detailed proof)
\begin{equation}
\eqalign{
\fl c_\nu(\mathcal{L}_n) = \sum_{i'} c_\nu(\mathcal{L}_\nu^{(i')}) \\
= \xi_\nu(\mathcal{L}_n) + \frac{1}{2}\sum_{\underset{\mu < \sigma}{\mu,\sigma}} \sum_{(\ell_\mu,\ell_\sigma)} \xi_{\ell_\mu \ell_\sigma}(\mathcal{L}_n) (\delta_{\mu,\nu}j(\ell_\nu) + \delta_{\sigma,\nu}j(\ell_\sigma))   \\
+ \frac{1}{2}(\delta_{\ell^{(1)},\uparrow_\nu}-\delta_{\ell^{(1)},\downarrow_\nu} + \delta_{\ell^{(n)},\uparrow_\nu}-\delta_{\ell^{(n)},\downarrow_\nu}) ,
}
\label{eq:generalizedcurr}
\end{equation}
with the last boundary term accounting for the first and last transitions in $\mathcal{L}_n$.

\subsection{Redundancy of mixed currents}

\label{sec:redundancy}

Given $M$ observable kinds, there are $(2M)^2$ possible pairs of subsequent transitions. $4M$ of them represents sequences of transitions of the same kind. Moreover, as mixed currents satisfy $\xi_{\ell_\nu\ell_\mu} = - \xi_{\bar{\ell}_\mu\bar{\ell}_\mu}$, we are left with $2M(M-1)$ mixed currents that are independent a priori. In this section, we apply KCL on the nodes of the transitions' space graph to further reduce the number of mixed currents, as not all of them are independent. KCL states that for each node $x \in X$ of a graph, the sum of all fluxes leaving $x$ and the fluxes entering $x$ is zero when the process is at stationarity. 

For simplicity we consider the case $M=2$ (see \fref{fig:model}(b)). KCL at nodes $\up_1$ and $\down_1$ reads
\begin{equation}
\eqalign{n_{\up_1\down_1} + n_{\up_1\up_2} + n_{\up_1\down_2} - n_{\down_1\up_1} - n_{\up_2\up_1} - n_{\down_2\up_1} = 0\\	
n_{\down_1\up_1} + n_{\down_1\up_2} + n_{\down_1\down_2} - n_{\up_1\down_1} - n_{\up_2\down_1} - n_{\down_2\down_1} = 0,
}
\end{equation}
when evaluated in closed sequences $\ell^{(1)} \to \cdots \ell^{(1)}$ of transitions. If we sum these two expressions, we find $\xi_{\up_1\up_2} + \xi_{\up_1\down_2} + \xi_{\down_1\up_2} + \xi_{\down_1\down_2} = 0$ (the same result would be found by employing KCL at $\up_2$ and $\down_2$). Since these mixed currents satisfy a linear relation, the number of independent mixed currents can be further reduced. However, in our following discussion we will allow ourselves some redundancy by considering all the $2M(M-1)$ mixed currents so that the FR can be written in a more symmetrical way. 

\subsection{Moment generating functions and FRs}

\label{sec:MGF}

In this section, we introduce the relation between the Moment Generating Function (MGF) for the joint statistics of observable currents and FRs. Given $M$ observable kinds, we denote by $\boldsymbol{c}$ the vector containing the $M$ total currents and by $\bxi$ the vector containing the $2M(M-1)$ mixed currents.

The MGF for the joint statistics $p_n(\boldsymbol{c},\bxi)$ of currents evaluated at $n$ occurrences of observable transitions is defined as
\begin{equation}
G_n(\boldsymbol{k},\bkappa) = \sum_{\boldsymbol{c}, \bxi \in \mathcal{F}_n} p_n(\boldsymbol{c},\bxi) \exp (\boldsymbol{k} \cdot \boldsymbol{c} + \bkappa \cdot \bxi ) ,
\label{eq:MGFdef}
\end{equation}
with $\mathcal{F}_n$ denoting the set of possible values that the currents can simultaneously take at a given $n$ (also called filtration). The counting fields $\boldsymbol{k}$ and $\bkappa$ are conjugated with the currents $\boldsymbol{c}$ and $\bxi$ respectively. If the joint statistics $p_n(\boldsymbol{c},\bxi)$ satisfies a detailed FR, asymptotically or at finite $n$ under the choice of a preferred initial probability in transition space (see \sref{sec:pref}), then
\begin{equation}
\ln \frac{p_n(\boldsymbol{c},\bxi)}{p_n( - \boldsymbol{c}, - \bxi)} = \boldsymbol{a} \cdot \boldsymbol{c} + \tilde{\balpha} \cdot \bxi.
\end{equation}
The symbol $\boldsymbol{a}$ denotes the vector of effective affinities conjugated to currents $\boldsymbol{c}$, each defined by Eq.\,\eqref{eq:affinitydef} and the symbol $\tilde{\balpha}$ the vector of ``shifted affinities'' conjugated with the mixed currents $\bxi$, whose exact expression will be derived in the following section. Those latter affinities are shifted with respect to the mixed affinities defined by Eq.\,\eqref{eq:aff}, and they are therefore called shifted mixed affinities. The reason why they appear will be clear in \sref{sec:FR} and is a consequence of Eq.\,\eqref{eq:generalizedcurr}. By plugging the expression above inside Eq.\,\eqref{eq:MGFdef}, we can express the detailed FR as a symmetry for the MGF
\begin{equation}
G_n(\boldsymbol{k},\bkappa) = G_n(-\boldsymbol{k} - \boldsymbol{a}, - \bkappa - \tilde{\balpha}) ,
\label{eq:MGFsym}
\end{equation}
which can be reached asymptotically at large $n$ or at all $n$ for a specific choice of initial probability. It is proven in \sref{app:genfuncurrents} that the MGF for the joint vector $(\boldsymbol{k},\bkappa)$ can be expressed as
\begin{equation}
G_n(\boldsymbol{k} , \bkappa) = \boldsymbol{1} \cdot \left[ P(\boldsymbol{k} , \bkappa) \right]^{n-1} E(\boldsymbol{k}) \boldsymbol{p}_1 ,
\label{eq:MGFalt}
\end{equation}
where we introduced the tilted trans-transition matrix
\begin{equation}
P(\boldsymbol{k},\bkappa)_{\ell_\nu\ell_\mu'}  =  \cases{P_{\ell_\nu\ell'_\nu} \exp (k_\nu j(\ell_\nu) ) &for  $\mu = \nu$ \\
P_{\ell_\nu\ell'_\mu} \exp (k_\nu j(\ell_\nu) + \kappa_{\ell_\nu\ell_\mu'}) &for   $\mu \neq \nu $, }
\label{eq:tiltedmatrixcurr}
\end{equation}
and the square matrix
\begin{equation}
E (\boldsymbol{x})_{\ell_\nu\ell'_\mu} =  \delta_{\ell_\nu,\ell_\mu'} \exp (x_\nu j(\ell_\nu) ) ,
\label{eq:matrixE}
\end{equation}
which only depends on quantities conjugated with the total currents and satisfies $J E (\boldsymbol{x}) = E(-\boldsymbol{x}) J$ (see \ref{app:0}) and $E (\boldsymbol{x})^{-1} =  E ( - \boldsymbol{x})$. The latter matrix is introduced to take into account the contribution to the total current carried by the first occurring transition.

A FR holds for the joint set $(\boldsymbol{k},\bkappa) $ if there exists a similarity transformation between $P(\boldsymbol{k},\bkappa)$ and $P( - \boldsymbol{k}- \boldsymbol{a}, - \bkappa - \tilde{\balpha})$. In particular, a FR holds if it is possible to find a square matrix $D$ such that
\begin{equation}
P(\boldsymbol{k},\bkappa)  =  E (\boldsymbol{k})  D^{-1} P(- \boldsymbol{a} - \boldsymbol{k},-\tilde{\balpha} - \bkappa)^\perp D E(- \boldsymbol{k}) . 
\label{eq:symcurr}
\end{equation}
The expression above represents indeed a similarity transformation as $J = J^{-1}$. Moreover, the use of the block-antitransposition Eq.\,\eqref{eq:antitranspdef} is convenient as $D$ can be found to be a diagonal matrix and it will also be used in \sref{sec:pref} to derive the preferred initial probabilities in transition space.

As briefly mentioned, the shifted affinities $\tilde{\balpha}$ were introduced as it is not possible to verify Eq.\,\eqref{eq:symcurr} with the use of the mixed affinities Eq.\,\eqref{eq:aff}. Thus, to proceed in the derivation of the FR for currents and mixed currents, we first need to derive an expression for the shifted affinities $\tilde{\balpha}$. This is done at the beginning of the next section by considering the log-ratio of path probabilities Eqs.\,\eqref{eq:pathFWD} and \eqref{eq:pathBWD}.

As a final remark before concluding this section, the MGF for a set of currents not including the mixed currents is obtained by setting $\boldsymbol{\kappa} = \boldsymbol{0}$, with $\boldsymbol{0}$ denoting the null vector.

\section{Results}

\label{sec:results}

\subsection{Fluctuation Relation}

\label{sec:FR}

In this section, we prove the central result of this paper, a FR for the joint set $\lbrace c_\nu \rbrace \cup \lbrace \xi_{\ell_\nu\ell_\mu} \rbrace$ of total currents and mixed currents. The proof is based on the symmetry Eq.\,\eqref{eq:MGFsym} for the MGF for the joint vector of currents $(\boldsymbol{c},\bxi)$, and relies in the correct identification of the mixed affinities to be conjugated with the mixed currents $\bxi$. We first address this point with the following consideration.

The logarithm of the ratio of Eq.\,\eqref{eq:pathFWD} and Eq.\,\eqref{eq:pathBWD} reads
\begin{equation}
\fl \ln \frac{p(\mathcal{L}_n)}{p(\bar{\mathcal{L}}_n)} = \sum_{\nu = 1}^M a_\nu \xi_\nu(\mathcal{L}_n) + \sum_{\underset{\nu < \mu}{\nu,\mu}} \sum_{(\ell_\nu,\ell_\mu)} \alpha_{\ell_\nu\ell_\mu}\xi_{\ell_\nu\ell_\mu}(\mathcal{L}_n) + u (\bar{\ell}^{(n)}) - u (\ell^{(1)}) ,
\end{equation}
with $u(\ell) = - \ln p_1(\ell)$ a potential. The expression above is written in terms of loop currents and mixed currents, which are respectively conjugated to the effective affinities Eq.\,\eqref{eq:affinitydef} and the mixed affinities Eq.\,\eqref{eq:aff}. We then employ Eq.\,\eqref{eq:generalizedcurr} to write the expression above in terms of the total currents. We obtain (see \ref{app:plug})
\begin{equation}
\fl \ln \frac{p(\mathcal{L}_n)}{p(\bar{\mathcal{L}}_n)} = \sum_{\nu = 1}^M a_\nu c_\nu(\mathcal{L}_n) + \sum_{\underset{\nu < \mu}{\nu,\mu}} \sum_{(\ell_\nu,\ell_\mu)} \tilde{\alpha}_{\ell_\nu\ell_\mu}\xi_{\ell_\nu\ell_\mu}(\mathcal{L}_n) + \tilde{u} (\bar{\ell}^{(n)}) - \tilde{u} (\ell^{(1)})
\label{eq:ratiocurrents}
\end{equation}
where we introduce new mixed affinities that are shifted with respect to Eq.\,\eqref{eq:aff} as
\begin{equation}
\tilde{\alpha}_{\ell_\nu\ell_\mu} = \alpha_{\ell_\nu \ell_\mu} - \frac{1}{2}(a_\nu j(\ell_\nu) + a_\mu j(\ell_\mu)),
\label{eq:tiltedcurredaff}
\end{equation}
called shifted mixed affinities. The boundary term is also modified as
\begin{equation}
\tilde{u} (\ell) = u(\ell) + \frac{1}{2}\left(\sum_{\nu} \sum_{\ell_\nu} a_\nu j(\ell_\nu)\delta_{\ell,\ell_\nu}\right) + v
\label{eq:tiltedcurredpot}
\end{equation}
with $v$ an arbitrary constant.

It is now possible to verify the symmetry Eq.\,\eqref{eq:MGFsym} for the MGF by finding a solution to the relation Eq.\,\eqref{eq:symcurr} by conjugating the total currents $\boldsymbol{c}$ with $\boldsymbol{a}$ and the mixed currents with the shifted mixed affinities $\tilde{\balpha}$ defined by Eq.\,\eqref{eq:tiltedcurredaff}. A diagonal matrix $D$ satisfying Eq.\,\eqref{eq:symcurr} is (see \ref{app:3symmetry} for a proof) is
\begin{equation}
D_{\ell_\nu \ell_\mu'} \propto \exp \left(-\frac{a_\nu}{2} j(\ell_\nu) \right) \delta_{\ell_\nu,\ell_\mu'}
\label{eq:diagmatrix}
\end{equation}
and thus the symmetry Eq.\,\eqref{eq:MGFsym} holds asymptotically for the joint set of total currents and mixed currents, thus satisfying a FR. The FR is written in the detailed form as
\begin{equation}
\fl \ln \frac{p(\lbrace c_{\nu} \rbrace, \lbrace \xi_{\ell_\nu\ell_\mu} \rbrace)}{p(\lbrace - c_{\nu} \rbrace, \lbrace - \xi_{\ell_\nu\ell_\mu} \rbrace)} = \sum_{\nu=1}^M a_\nu c_\nu + \sum_{\underset{\mu > \nu}{\nu,\mu}} \sum_{(\ell_\nu,\ell_\mu)} \tilde{\alpha}_{\ell_\nu\ell_\mu} \xi_{\ell_\nu\ell_\mu} +\tilde{u} (\bar{\ell}^{(n)}) - \tilde{u} (\ell^{(1)}) ,
\label{eq:asymptoticFR}
\end{equation}
by keeping the explicit dependence on the boundary potential. Notice that in the presence of a single observable current, the expression above reduces to the result found in Ref.\,\cite{beatofacurrent}.

\subsection{Complete sets of currents}

\label{sec:completecurr}

In this section, we show that Eq.\,\eqref{eq:asymptoticFR} reduces to the FR for a complete set of currents \cite{Andrieux} where now the observation process is arrested after the occurrence of $n$ observable transitions instead of the clock time $t$. In graph-theoretical terminology, as explained in \cite{Andrieux,schnakenberg}, the set of observed edges $\mathcal{O}$ is complete if $E \setminus \lbrace \boldsymbol{e}_\nu \rbrace$ is a spanning tree $\mathscr{T}$, which contains no cycles. Reinsertion of a chord $\boldsymbol{e}_\nu$ identifies a cycle after cancelation of remaining branches that do not belong to the cycle. In this framework, we show that observations along chords $\boldsymbol{e}_\nu \in E$ satisfying the conditions above lead to a FR for the currents $\lbrace c_\nu \rbrace$ without the need to include mixed currents.

An important feature of complete sets is that every pair of states $x,y \in X$ is connected by a unique path on the spanning tree $\mathscr{T}$, and since $\mathscr{T}$ contains no cycles, each sequence $x_1 \to x_2 \to \cdots \to x_1$ with transitions occurring in $\mathscr{T}$ is reversible, in the sense that it does not produce entropy. This fact can be expressed in terms of the transition probabilities $\pi(x|y)$, defined by Eq.\,\eqref{eq:transitionprobabilities}, of the embedded state space chain
\begin{equation}
\frac{\pi(x_1|x_{n-1})\cdots \pi(x_2|x_1)}{\pi(x_1|x_2)\cdots \pi(x_{n-1}|x_1)} = 1 .
\label{eq:statesrev}
\end{equation}
In our transition-based formalism, since the spanning tree $\mathscr{T}$ contains no cycles, the trans-transition probabilities can be factorized as $p(\ell|\ell') = \sigma(\ell|\ell') \pi(\Gamma(\mathtt{t}(\ell') \to \mathtt{s}(\ell))) \pi(\ell)$ (see \ref{app:hiddenpathcomplete}), where $\Gamma(\mathtt{t}(\ell') \to \mathtt{s}(\ell))$ is the path along a spanning tree connecting $\mathtt{t}(\ell')$ to $\mathtt{s}(\ell)$ with the smallest number of edges (also known as backbone in Section IV of Ref.\,\cite{harunari}), and $\sigma(\ell|\ell')$ denotes the multiplicative contribution accounting for futile excursions out of such path but still on the tree.
Moreover, $\pi(\ell) = \pi(\mathtt{t}(\ell) | \mathtt{s}(\ell))$ is a simplified notation for the transition probability of $\ell$ conditioned on its source state. Notice that for simplicity, we also use the symbol $\pi$ as an operator $\pi(\Gamma)$ that associates with a generic trajectory $\Gamma$ in states' space the product $\pi(x_n|x_{n-1}) \cdots \pi(x_2|x_1)$ of the transition probabilities along the path from state $x_1$ to $x_n$, whereas if $\Gamma$ consists in a single jump, it denotes the probability associated with that jump. For the time-reversed sequence of visible transitions, the trans-transition probability $p(\bar{\ell}'|\bar{\ell}) = \sigma(\bar{\ell}'|\bar{\ell}) \pi(\Gamma(\mathtt{t}(\bar{\ell}) \to \mathtt{s}(\bar{\ell}'))) \pi(\bar{\ell}')$ contains the path in the opposite direction, $\Gamma(\mathtt{s}(\ell) \to \mathtt{t}(\ell'))$, since $\mathtt{s}(\ell) = \mathtt{t}(\bar{\ell})$. The contribution from futile excursions is symmetric under time-reversal $\sigma(\ell|\ell') = \sigma(\bar{\ell}'|\bar{\ell})$ (see \ref{app:hiddenpathcomplete} for a proof) since they occur with the same probability as the original path. By consequence, for all $\ell_\nu, \ell_\mu$,
\begin{equation}
\frac{p(\ell_\nu|\ell_\mu)}{p(\bar{\ell}_\mu |\bar{\ell}_\nu)}  = \frac{\pi(\ell_\nu)}{\pi(\bar{\ell}_\mu)} \frac{\pi(\Gamma (\mathtt{t}(\ell_\mu) \to \mathtt{s}(\ell_\nu)))}{\pi(\Gamma(\mathtt{s}(\ell_\nu) \to \mathtt{t}(\ell_\mu)))}.
\label{eq:ratiotimerev}
\end{equation}
Finally, Eq.\,\eqref{eq:statesrev} can be expressed in terms of trans-transition probabilities as
\begin{equation}
\frac{p(\ell^{(1)}|\bar{\ell}^{(n-1)})}{p(\ell^{(n-1)}|\bar{\ell}^{(1)})} \cdots \frac{p(\ell^{(2)}|\bar{\ell}^{(1)})}{p(\ell^{(1)}|\bar{\ell}^{(2)})}  = 1
\label{eq:transrev}
\end{equation}
for each cyclic sequence of $n$ observable transitions $\ell^{(1)}\to \ell^{(2)}\to \cdots \to \ell^{(n-1)} \to \ell^{(1)}$. We call this latter the hidden equilibrium condition. The expressions Eqs.\,\eqref{eq:statesrev} and \eqref{eq:transrev} are equivalent as by virtue of Eq.\,\eqref{eq:ratiotimerev}, the surviving terms in Eq.\,\eqref{eq:transrev} are transition probabilities between states along the spanning tree $\mathscr{T}$. Notice that the hidden equilibrium condition expressed by Eqs.\,\eqref{eq:statesrev} and \eqref{eq:transrev} is also satisfied in the case where the hidden subgraph contains futile cycles, i.e. cycles with vanishing affinity.

Before proving the FR for a complete set of currents, without considering mixed currents, we use Eq.\,\eqref{eq:transrev} to provide two properties that are satisfied by the mixed affinities Eq.\,\eqref{eq:aff} when the paths in the hidden network are at equilibrium.

First, we consider the nontrivial cyclic sequence in transition space $\ell_\nu \to \ell_\rho \to \ell_\mu \to \ell_\nu$ in a process with three observable kinds $\nu\neq \rho\neq \mu$, where no two consecutive transitions occur along the same edge. Then Eq.\,\eqref{eq:transrev} provides
\begin{equation}
\alpha_{\ell_\nu\bar{\ell}_\rho} + \alpha_{\ell_\rho\bar{\ell}_\mu} = \alpha_{\ell_\nu\bar{\ell}_\mu}
\label{eq:affinitescomplete1}
\end{equation}
for all $\nu$, $\rho$, $\mu$ with $\nu \neq \rho\neq \mu$. If only two kinds are observable, the expression above reduces to the antisymmetric property $\alpha_{\ell_\nu\ell_\mu'} = -\alpha_{\bar{\ell}_\mu'\bar{\ell}_\nu}$ with respect to the notion of time reversal explained in \sref{sec:time-reversal}. The second property is found by considering sequences such as $\ell_\nu \to \ell_\mu \to \bar{\ell}_\mu \to \ell_\nu$, yielding
\begin{equation}
\alpha_{\ell_\nu\bar{\ell}_\mu} = \alpha_{\ell_\nu\ell_\mu} - a_\mu j(\ell_\mu),
\label{eq:affinitescomplete2}
\end{equation}
which holds for all $\nu, \mu$ with $\nu\neq \mu$. It can be checked that every closed sequence $\ell^{(1)} \to \cdots \to \ell^{(1)}$ satisfies Eq.\,\eqref{eq:transrev} by combining Eqs.\,\eqref{eq:affinitescomplete1} and \eqref{eq:affinitescomplete2}.

We now prove the FR for complete sets of currents by considering the symmetry Eq.\,\eqref{eq:symcurr} for the MGF Eq.\,\eqref{eq:MGFdef}. Specifically, we are interested in the statistics of the $M$ observable total currents where we exclude the counting of mixed occurrences $\ell_\mu \to \ell_\nu$ with $\nu \neq \mu$. Thus, the tilted matrix $P(\boldsymbol{k},\boldsymbol{0})$ now depends only on the counting fields $\boldsymbol{k}$, each of its components $k_\nu$ being conjugated with the total current $c_\nu$. Hence, a matrix element in Eq.\,\eqref{eq:symcurr} is explicitly
\begin{equation}
p(\ell_\nu|\ell_\mu ') \exp (k_\mu j (\ell_\mu ')) = \frac{d(\ell_\mu ')}{d(\ell_\nu)} p(\bar{\ell}_\mu ' |\bar{\ell}_\nu ) \exp((a_\mu + k_\mu) j(\ell_\mu '))
\label{eq:matrixelementcomplete}
\end{equation}
with $a_\nu$ the effective affinity Eq.\,\eqref{eq:affinitydef} conjugated to the total current $c_\nu$ and where $d(\ell_\nu)$ denotes the element of the diagonal matrix $D$ corresponding to transition $\ell_\nu$. Selecting entries with $\ell_\mu' = \bar{\ell}_\nu$, Eq.\,\eqref{eq:matrixelementcomplete} provides
\begin{equation}
\frac{d(\ell_\nu)}{d(\bar{\ell}_\nu)} = \exp (-a_\nu j(\ell_\nu)),
\label{eq:prop1}
\end{equation}
that once plugged back inside Eq.\,\eqref{eq:matrixelementcomplete} gives for $\mu \neq \nu$
\begin{equation}
\frac{d(\ell_\nu)}{d(\ell_\mu)} = \exp (-\alpha_{\ell_\nu\bar{\ell}_\mu}) .
\label{eq:prop2}
\end{equation}
The conditions {Eq.\,}\eqref{eq:prop1} and {Eq.\,}\eqref{eq:prop2} must be satisfied by the matrix elements of $D$ so that the joint probability $p_n(c_1,\dots, c_M)$ satisfies a FR. It is now very easy to show that {Eq.\,}\eqref{eq:prop1} and {Eq.\,}\eqref{eq:prop2} are compatible with the requirements {Eq.\,}\eqref{eq:affinitescomplete1} and {Eq.\,}\eqref{eq:affinitescomplete2} for complete sets of currents. In fact, by plugging Eq.\,\eqref{eq:prop2} into Eq.\,\eqref{eq:matrixelementcomplete} for $\mu\neq \nu$ and using the definition Eq.\,\eqref{eq:aff} one finds the relation Eq.\,\eqref{eq:affinitescomplete2} for the mixed affinities. By considering {Eq.\,}\eqref{eq:prop2}, since it has to hold for all pairs of kinds $\nu$, $\mu$ with $\nu\neq\mu$
\begin{equation}
\frac{d(\ell_\nu)}{d(\ell_\mu)} = \frac{d(\ell_\nu	)}{d(\ell_\rho)} \frac{d(\ell_\rho)}{d(\ell_\mu)} = \exp(- a_{\ell_\nu\bar{\ell}_\mu}) = \exp(- a_{\ell_\nu\bar{\ell}_\rho} - a_{\ell_\rho\bar{\ell}_\mu}) ,
\end{equation}
which is equivalent to Eq.\,\eqref{eq:affinitescomplete1}. The conditions Eqs.\,\eqref{eq:prop1} and \eqref{eq:prop2} are only verified if the mixed affinities satisfy Eqs.\,\eqref{eq:affinitescomplete1} and \eqref{eq:affinitescomplete2}. These properties emerge from the hidden equilibrium condition Eq.\,\eqref{eq:transrev} which is always satisfied when the set of observable transitions is complete. We thus conclude that the joint probability $p_n(c_1,\dots, c_M)$ evaluated for the complete set $\lbrace c_\nu \rbrace$ satisfies a FR where the mixed currents do not appear, as its MGF satisfies the symmetry Eq.\,\eqref{eq:MGFsym} up to boundary terms, which corresponds to
\begin{equation}
\ln \frac{p_n(c_1,\dots,c_M)}{p_n (-c_1,\dots,-c_M)} = \sum_{\nu = 1}^M a_\nu c_\nu + \Delta\tilde{\tilde{u}}(\bar{\ell}^{(n)}, \ell^{(1)}) ,
\label{eq:completesetFR}
\end{equation}
with $\tilde{\tilde{u}}$ a boundary potential.

In the next section, we provide additional considerations on the case of complete sets of currents. The properties Eqs.\,\eqref{eq:affinitescomplete1} and \eqref{eq:affinitescomplete2} for the mixed affinities affects, in fact, the shifted affinities Eq.\,\eqref{eq:tiltedcurredaff}. As a result, it is possible to associate the trans-transitions between different kinds, regardless of their orientation, with a process in the space of kinds $\lbrace \nu \rbrace_{\nu = 1}^M$ with vanishing cycle affinities.

\subsubsection{Equilibrium in the space of kinds}

\label{sec:typeseq}

For a physical interpretation of the result Eq.\,\eqref{eq:completesetFR}, we consider the mixed affinities $\alpha_{\ell_\nu\ell_\mu}$ defined by Eq.\,\eqref{eq:aff}. Consequently to Eq.\,\eqref{eq:ratiotimerev}, it can be shown (\ref{app:6}) that 
\begin{equation}
\alpha_{\ell_\nu\ell_\mu} = \ln\frac{p(\ell_\nu|\ell_\mu)}{p(\bar{\ell}_\mu|\bar{\ell}_\nu)} = \ln \left[ \omega_{\nu\mu}\frac{p(\ell_\nu|\ell_\nu)}{p(\bar{\ell}_\mu|\bar{\ell}_\mu)} \right]
\label{eq:coeffs}
\end{equation}
for each pair of transitions $(\ell_\nu,\ell_\mu)$, compatible with Eq.\,\eqref{eq:affinitescomplete2}. The coefficients $\omega_{\nu\mu}$ are independent of the choices of $\ell_\nu$ and $\ell_\mu$ and only depend on their kinds $\nu$ and $\mu$, and they are given by
\begin{equation}
 \omega_{\nu\mu} := \frac{p(\ell_\nu|\ell_\mu)}{p(\bar{\ell}_\mu|\bar{\ell}_\nu)} \frac{p(\bar{\ell}_\mu|\bar{\ell}_\mu)}{p(\ell_\nu|\ell_\nu)}.
\end{equation}
Moreover, by using the definition Eq.\,\eqref{eq:tiltedcurredaff} for the shifted mixed affinities $\tilde{\alpha}_{\ell_\nu\ell_\mu}$ appearing in the FR Eq.\,\eqref{eq:asymptoticFR} one obtains
\begin{equation}
\tilde{\alpha}_{\nu\mu} := \tilde{\alpha}_{\ell_\nu\ell_\mu} = \ln \left[ \omega_{\nu\mu}\left(\frac{p(\up_\nu|\up_\nu) p(\down_\nu|\down_\nu)}{p(\up_\mu|\up_\mu) p(\down_\mu|\down_\mu)}\right)^{\frac{1}{2}} \right]
\label{eq:tiltedcurredaffcomplete}
\end{equation}
now depending only on the kinds $\nu$ and $\mu$. In fact, each pair of transitions $\ell_\nu \in \lbrace \up_\nu,\down_\nu \rbrace$ and $\ell_\mu \in \lbrace \up_\mu, \down_\mu \rbrace$ provides the same value of Eq.\,\eqref{eq:tiltedcurredaffcomplete}. Also notice that for complete sets of currents, the shifted mixed affinities Eq.\,\eqref{eq:tiltedcurredaffcomplete} become antisymmetric with respect to $\nu$ and $\mu$, being $\tilde{\alpha}_{\nu\mu} = - \tilde{\alpha}_{\mu\nu}$ (the same does not happen to the unshifted affinities Eq.\,\eqref{eq:coeffs}). Following Eq.\,\eqref{eq:tiltedcurredaffcomplete} the mixed contributions to the FR Eq.\,\eqref{eq:asymptoticFR} can be rewritten in terms of the affinities $\tilde{\alpha}_{\nu\mu}$ and the intertype currents $\xi_{\nu\mu} = \sum_{\ell_\nu\ell_\mu} \xi_{\ell_\nu\ell_\mu}$, resulting in the fact that the second term on the RHS of Eq.\,\eqref{eq:asymptoticFR} is bounded and thus does not appear in the asymptotic FR Eq.\,\eqref{eq:completesetFR} for complete sets.

In the simple case $M=2$, the mixed contribution to the FR Eq.\,\eqref{eq:asymptoticFR} is
\begin{equation}
\tilde{\alpha}_{12} \left( \xi_{\up_1\up_2} + \xi_{\up_1\down_2} + \xi_{\down_1\up_2} + \xi_{\down_1\down_2} \right) = \tilde{\alpha}_{12} \xi_{12}.
\label{eq:mixedterm2trans}
\end{equation}
In this case, $\xi_{12}$ only takes values $\pm 1$ and 0 regardless of the length of the full sequence $\mathcal{L}_n$, since it is only determined by the kinds of the first and last transition. \Eref{eq:mixedterm2trans} can then be incorporated in the boundary potential $\Delta \tilde{u}$, which does not contribute in the limit $n\to \infty$ (but could be absorbed by the choice of a suitable initial distribution).

\begin{figure}
\centering
\includegraphics[scale=1]{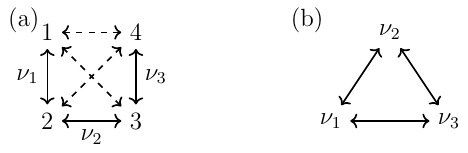}
\caption{(a) An example of a graph with three observable transitions $\nu_1$, $\nu_2$ and $\nu_3$ forming a complete set. (b) The process between kinds of transitions, i.e. transitions occurring on different edges regardless their orientation, can be represented by a graph with nodes the transitions' kinds and with edges the transitions between different kinds. The number of independent cycles can be found by use of Schnakenberg cycle decomposition in this space, thus finding relations between mixed affinities which allows recovering of the FR \cite{Andrieux} for complete sets of currents.}
\label{fig:intertype}
\end{figure}

For $M>2$ we can establish similar relations between the shifted affinities, but they are not all independent. We then consider an alternative description based on occurrences of kinds rather than directed transitions (see Fig.\,\ref{fig:intertype}). In this sense, in a realization $\mathcal{L}_n$ of the process in the observable transition space, we are only interested in subsequent events occurring along different observable edges, regardless of their orientations being $\up_\nu$ or $\down_\nu$. In simple words, for complete sets of currents, the shifted affinities $\tilde{\alpha}_{\nu\mu}$ drive the transitions from snippets of kind $\mu$ to kind $\nu$ and viceversa, since $\tilde{\alpha}_{\nu\mu} = -\tilde{\alpha}_{\mu\nu}$. We define a process in the space of kinds $\lbrace \nu \rbrace_{\nu = 1}^M$ of observable transitions where transitions between kinds are driven by mixed affinities $\tilde{\alpha}_{\nu\mu}$. Each cycle in the process in the space of kinds has zero affinity, i.e. for all cyclic sequences of kinds $\nu_1 \to \nu_2 \to \cdots \to \nu_{m-1} \to \nu_1$ of any length $m \geq 2$
\begin{equation}
\fl \tilde{\alpha}_{\nu_1 \nu_2} +\tilde{\alpha}_{\nu_2 \nu_3} + \cdots + \tilde{\alpha}_{\nu_{m-2} \nu_{m-1}} + \tilde{\alpha}_{\nu_{m-1} \nu_1} = \ln \left( \omega_{\nu_1\nu_2} \cdots \omega_{\nu_{m-1}\nu_1} \right) = 0
\label{eq:equilibrium}
\end{equation}
if the observable transitions form a complete set, as it follows directly from Eq.\,\eqref{eq:transrev}. In this space, the relation above is equivalent to Kolmogorov's condition for equilibrium, and therefore it implies the existence of potentials $\upsilon_\nu$ each associated with a single kind $\nu$, such that the affinities $\tilde{\alpha}_{\nu\mu}$ can be expressed as the difference
\begin{equation}
\tilde{\alpha}_{\nu\mu} = \upsilon_\nu - \upsilon_\mu .
\label{eq:ansatz}
\end{equation}
We see in fact that Eq.\,\eqref{eq:ansatz} preserves the property Eq.\,\eqref{eq:equilibrium} and thus the composition rule Eq.\,\eqref{eq:affinitescomplete1}
\begin{equation}
\tilde{\alpha}_{\nu \rho} + \tilde{\alpha}_{\rho\mu} = \tilde{\alpha}_{\nu\mu}
\end{equation}
extended to the shifted affinities $\tilde{\alpha}_{\nu\mu}$ in the case of complete sets of currents. Hence, the mixed contribution in Eq.\,\eqref{eq:asymptoticFR} can be expressed in terms of the potentials $\upsilon_\nu$ as
\begin{equation}
\sum_{\underset{\mu>\nu}{\nu,\mu}}\tilde{\alpha}_{\nu\mu} \xi_{\nu\mu} = -\sum_\nu \upsilon_\nu \sum_{\mu \neq \nu} \xi_{\mu \nu}.
\label{eq:intertypenet}
\end{equation}
The sum $\sum_{\mu \neq \nu} \xi_{\mu \nu}$ is interpreted as the difference between the number of times the system leaves and reaches kind $\nu$ in a single realization $\mathcal{L}_\nu$ of the process in transition space. For cyclic sequences of kinds $\nu\to \cdots \to \nu$ it is then $\sum_{\mu\neq\nu} \xi_{\mu \nu} = 0$. Thus Eq.\,\eqref{eq:intertypenet} only depends on the kinds of the first and last transition. In particular, it is reabsorbed in the potential term Eq.\,\eqref{eq:tiltedcurredpot}, defining a new potential
\begin{equation}
\tilde{\tilde{u}} (\ell) = \tilde{\upsilon}(\ell) + \sum_{\nu}\sum_{\ell_\nu} \upsilon_\nu \delta_{\ell,\ell_\nu} .
\label{eq:tiltedtiltedpot}
\end{equation}
As the potential above is bounded, the first term in the RHS of Eq.\,\eqref{eq:asymptoticFR}, which contains the total currents, dominates at large $n$.

We conclude that for complete sets of currents the FR Eq.\,\eqref{eq:completesetFR} holds as a consequence of the reversibility of all closed paths on the spanning tree $\mathscr{T}$. In fact, this condition provides the symmetry Eq.\,\eqref{eq:symcurr} for the MGF and the fact that the intertype process is at equilibrium, as stated by Eq.\,\eqref{eq:equilibrium}.

\subsection{Transient FRs}

\label{sec:pref}

In this section, we extend the results Eq.\,\eqref{eq:asymptoticFR} and Eq.\,\eqref{eq:completesetFR} to the case where the observation process is stopped after a finite number $n$ of observable transitions. Notice that Eq.\,\eqref{eq:ratiocurrents} is already expressed at finite $n$, yet it contains the explicit dependence on a boundary term that contains the initial probabilities $p_1(\ell)$ for the first (or last) transition being $\ell$. An exact FR (without boundary terms) is obtained from Eq.\,\eqref{eq:ratiocurrents} when the boundary transitions can be marginalized, which can be done in general when realizations $\mathcal{L}_n$ are post-selected so that the boundary term $\Delta \tilde{u} := \tilde{u}(\bar{\ell}^{(n)}) - \tilde{u}(\ell^{(1)})$ vanishes (which is the case when $\bar{\ell}^{(n)} = \ell^{(1)}$).

For noncomplete sets of currents, we search for probabilities $p_1^*(\ell)$ such that the boundary term $\Delta \tilde{u} $ vanishes at all $n$, without the need to post-select sequences $\mathcal{L}_n$. For complete sets, we must further impose that the combined effects of $\Delta \tilde{u}$ in Eq.\,\eqref{eq:ratiocurrents} and the bounded term containing the mixed currents (see \sref{sec:typeseq}) vanish.

The task of finding a preferred initial probability $\boldsymbol{p}_1^*$ (in vector form) is achieved by comparing both sides in Eq.\,\eqref{eq:MGFsym} when the MGF is written in the form  Eq.\,\eqref{eq:MGFalt} and by use of Eq.\,\eqref{eq:symcurr}. As discussed previously, the existence of a real diagonal matrix $D$ satisfying Eq.\,\eqref{eq:symcurr} is enough to state that a FR is satisfied by the observed currents. Denoting with $\boldsymbol{k}$ the vector containing the counting fields $k_\nu$ conjugated to the total currents $c_\nu$ and with $\bkappa$ the vector containing the counting fields $\kappa_{\ell_\nu\ell_\mu}$ conjugated to the mixed currents $\xi_{\ell_\nu\ell_\mu}$, the choice
\begin{equation}
\boldsymbol{p}_1^*\propto D^{-1}\boldsymbol{1}
\label{eq:preferreddef}
\end{equation}
for the initial distribution provides the symmetry
\begin{equation}
G_n(\boldsymbol{k},\bkappa) = G_n(-\boldsymbol{a} - \boldsymbol{k}, -\tilde{\balpha} - \bkappa)
\label{eq:MGFfinite}
\end{equation}
at all times $n$, with $\boldsymbol{a}$ the vector containing the effective affinities $a_\nu$ driving the total currents $c_\nu$ and $\tilde{\balpha}$ the vector containing the shifted mixed affinities $\tilde{\alpha}_{\ell_\nu\ell_\mu}$. 

The left-hand side of Eq.\,\eqref{eq:MGFfinite} is in fact
\begin{equation}
G_n(\boldsymbol{k},\bkappa) =  \boldsymbol{1} \cdot E(\boldsymbol{k}) D^{-1} \left[P(-\boldsymbol{a} - \boldsymbol{k}, -\tilde{\balpha} - \bkappa)^{\perp} \right]^{n-1} D \boldsymbol{p}_1 ,
\label{eq:proofFR}
\end{equation}
where we used the relation Eq.\,\eqref{eq:symcurr}. Since the expression above is a scalar product, we can transpose all the quantities obtaining
\begin{equation}
G_n(\boldsymbol{k},\bkappa)  = \boldsymbol{p}_1 \cdot D J \left[P(-\boldsymbol{a} - \boldsymbol{k}, -\tilde{\balpha} - \bkappa) \right]^{n-1} E (-\boldsymbol{k})  J D^{-1}  \boldsymbol{1}
\label{eq:leftside}
\end{equation}
where we also used that $(P^{\perp})^\top = J P J$ and $J E (\boldsymbol{k}) = E (-\boldsymbol{k}) J$ (see \ref{app:0}).

The RHS of Eq.\,\eqref{eq:MGFfinite} is given by
\begin{equation}
\fl G_n(-\boldsymbol{a} - \boldsymbol{k}, -\tilde{\balpha} - \bkappa) = \boldsymbol{1}\cdot \left[ P(-\boldsymbol{a} - \boldsymbol{k}, -\tilde{\balpha} - \bkappa) \right]^{n-1} E (-\boldsymbol{k})E (\boldsymbol{ -\boldsymbol{a}}) \boldsymbol{p}_1
\label{eq:rightside}
\end{equation}
By comparing Eqs. Eq.\,\eqref{eq:leftside} and Eq.\,\eqref{eq:rightside} we finally see that by taking $\boldsymbol{p}_1^* = D^{-1}\boldsymbol{1}$ then the symmetry Eq.\,\eqref{eq:MGFfinite} holds at all $n$. In fact, for this choice of $\boldsymbol{p}_1$ the following identity
\begin{equation}
E (-\boldsymbol{a}) \boldsymbol{p}_1^* = J \boldsymbol{p}_1^*
\end{equation}
is also verified (see \ref{app:4}).

We now consider the case of noncomplete sets of currents, where observation of all mixed currents to complement the $\lbrace c_\nu\rbrace$ is necessary to achieve the FR Eq.\,\eqref{eq:asymptoticFR}. For this case Eq.\,\eqref{eq:preferreddef} provides after normalization and a few manipulations (\ref{app:3preferred})
\begin{equation}
p_{1,{\rm nc}}^*(\ell_\nu) = \frac{\exp \left(\frac{1}{2}a_\nu j(\ell_\nu)\right)}{2 \sum_\mu \cosh\left( \frac{1}{2} a_\mu \right)}.
\label{eq:preferredprobability}
\end{equation}

For complete sets of currents we use the solution of $D$ based on Eq.\,\eqref{eq:prop1} and Eq.\,\eqref{eq:prop2}. As explained in \ref{app:3complete} we can use the ansatz Eq.\,\eqref{eq:ansatz} to write the shifted mixed affinities $\tilde{\alpha}_{\nu\mu}$ in terms of the differences of potentials $\upsilon_\nu$ and $\upsilon_\mu$. With this choice, by repeating the same scheme used for Eq.\,\eqref{eq:preferredprobability} one finds
\begin{equation}
p_{1,{\rm c}}^*(\ell_\nu) = \frac{\exp \left(\frac{1}{2}a_\nu j(\ell_\nu) + \upsilon_\nu\right)}{2 \sum_\mu \exp(\upsilon_\mu) \cosh\left( \frac{1}{2} a_\mu \right)}.
\label{eq:preferredcomplete}
\end{equation}

These preferred distributions ensure that the FRs Eqs.\,\eqref{eq:asymptoticFR} and \eqref{eq:completesetFR} are satisfied of all times since Eq.\,\eqref{eq:preferredprobability} cancel the potentials $\tilde{u} (\ell),\ \forall\ell$ Eq.\,\eqref{eq:tiltedcurredpot} and Eq.\,\eqref{eq:preferredcomplete} cancels $\tilde{\tilde{u}} (\ell),\ \forall\ell$ Eq.\,\eqref{eq:tiltedtiltedpot}, if the arbitrary constant $v$ in Eq.\,\eqref{eq:tiltedcurredpot} is chosen as the normalization factor for the initial distribution.

\section{Discussion and conclusions}

\label{sec:conclusions}

The convention of timekeeping by ticks of a clock is a social construct, while time's flowing direction is not. We have shown that the FR---a quantifier of distinguishability between trajectories and their time reversals---can be recovered in the case of hidden currents if the notion of time is related to the observable activity [see Eq.\,\eqref{eq:asymptoticFR}], highlighting how key thermodynamic properties are preserved by instrinsic definitions of random time \cite{PhysRevLett.119.140604}. Our result bridges the gap between the case of a single observable current \cite{beatofacurrent} and a complete set \cite{Andrieux_2007} by the vanishing contribution of mixed currents [see Eq.\,\eqref{eq:completesetFR}].

\begin{figure}
\centering
\includegraphics[width=0.9\textwidth]{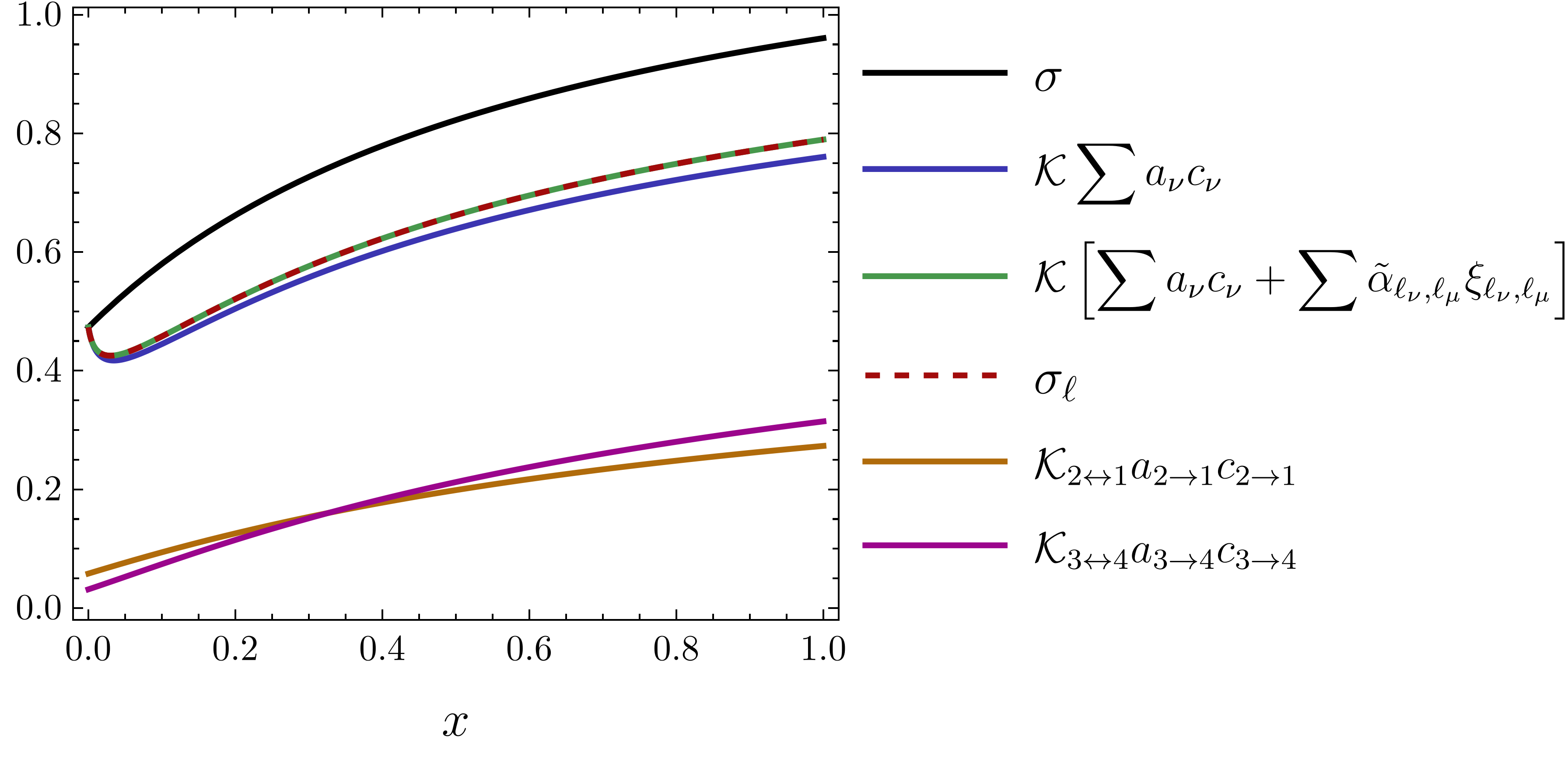}
\caption{For the system depicted in \fref{fig:model} with $x = r_{42} = 5\,r_{24}$ and all other rates randomized between 0 and 1, we calculate the entropy production rate (black) and the RHS of Eq.\,\eqref{eq:asymptoticFR} in the following scenarios: observation of $1 \leftrightarrow 2$ (orange), observation of $3 \leftrightarrow 4$ (purple), observation of both (green), and observation of both without the mixed terms (blue). The red dashed line represents the contribution to entropy production by the sequences of transitions given in \cite{harunari} and perfectly overlaps with the green line. Finally, $x = 0$ corresponds to eliminating one of the cycles, rendering the set $\{ 1 \leftrightarrow 2, 3 \leftrightarrow 4 \}$ complete. The observable traffic rate $\mathcal{K} = \mathcal{K}_{2\leftrightarrow 1} + \mathcal{K}_{3\leftrightarrow 4}$ is required to change from unit transitions to time, and is defined as the number of observable transitions over the respective edge divided by the time duration.}
\label{fig:entropy}
\end{figure}

These arrow of time quantifications by means of FRs are related to the notion of dissipation, measured by the entropy production. In its usual version, the RHS is recognized as the entropy production rate and, as a direct consequence of the FR, is nonnegative on average, constituting the nonequilibrium second law. The RHSs of Eqs.\,\eqref{eq:asymptoticFR} and \eqref{eq:completesetFR} are nevertheless measures of the observable dissipation, which can be understood by the fact that their averages represent the relative entropy between the probability of observable currents and their time reversal counterpart. As shown in \cite{harunari, vandermeer_2022}, the product of the current and the effective affinity bounds the entropy production rate from below; analogously, the RHS of the FR derived in \cite{beatofacurrent} provides the same bound when a single current is observed, and the inclusion of mixed currents/affinities improves it when more currents are observed.
In \Fref{fig:entropy}, we illustrate how these quantities provide a lower bound for the entropy production rate. We consider a four-state process where the currents flowing through $1 \leftrightarrow 2$ and/or $3\leftrightarrow 4$ are observed. The complete set is formed by three currents defined over different cycles, thus observing two transitions leads to a lower bound. The inclusion of mixed terms makes it tighter and matches the value of $\sigma_\ell$ from \cite{harunari} (see blue and green lines), showcasing how $\sigma_\ell$ already encompasses this cross-information. When only one current is observed, the bound gets looser (see orange and purple lines). For $x = 0$, the two observed currents form a complete set, and we see that \textit{(i)} mixed terms stop contributing due to the collapse of blue and green lines; \textit{(ii)} the exact entropy production is obtained due to the collapse with the black line. It would be interesting to explore in more detail the role of mixed currents in the growing discourse of entropy production estimation.

Importantly, we point out that mixed currents are not an additional ingredient inserted into the theory, they emerge as relevant observables in the scenario of partial observation of currents. The probability of full trajectories and their reverse satisfies a fluctuation relation, which can be decomposed into terms containing currents and their correlations. The mixed currents capture these correlations and, as we have shown, their contribution is absent in the complete set case (see point \textit{(i)} of the paragraph above). However, there is no a priori reason to believe they would not contribute when a non-complete set is considered, and they do.

The initial distribution that satisfies the symmetry Eq.\,\eqref{eq:MGFsym} at all $n$, for both non-complete [Eq.\,\eqref{eq:preferredprobability}] and complete [Eq.\,\eqref{eq:preferredcomplete}] sets, ensure the FR even at small recorded activities. From a practical viewpoint, a transient FR allows the estimation of effective affinities from short trajectories, thus circumventing the limitations imposed by sampling events in the tails of distributions, which become increasingly rare as $n$ increases. However, an operational interpretation of the preferred distribution is still missing.

\begin{figure}
\centering
\includegraphics[height=.32\textheight]{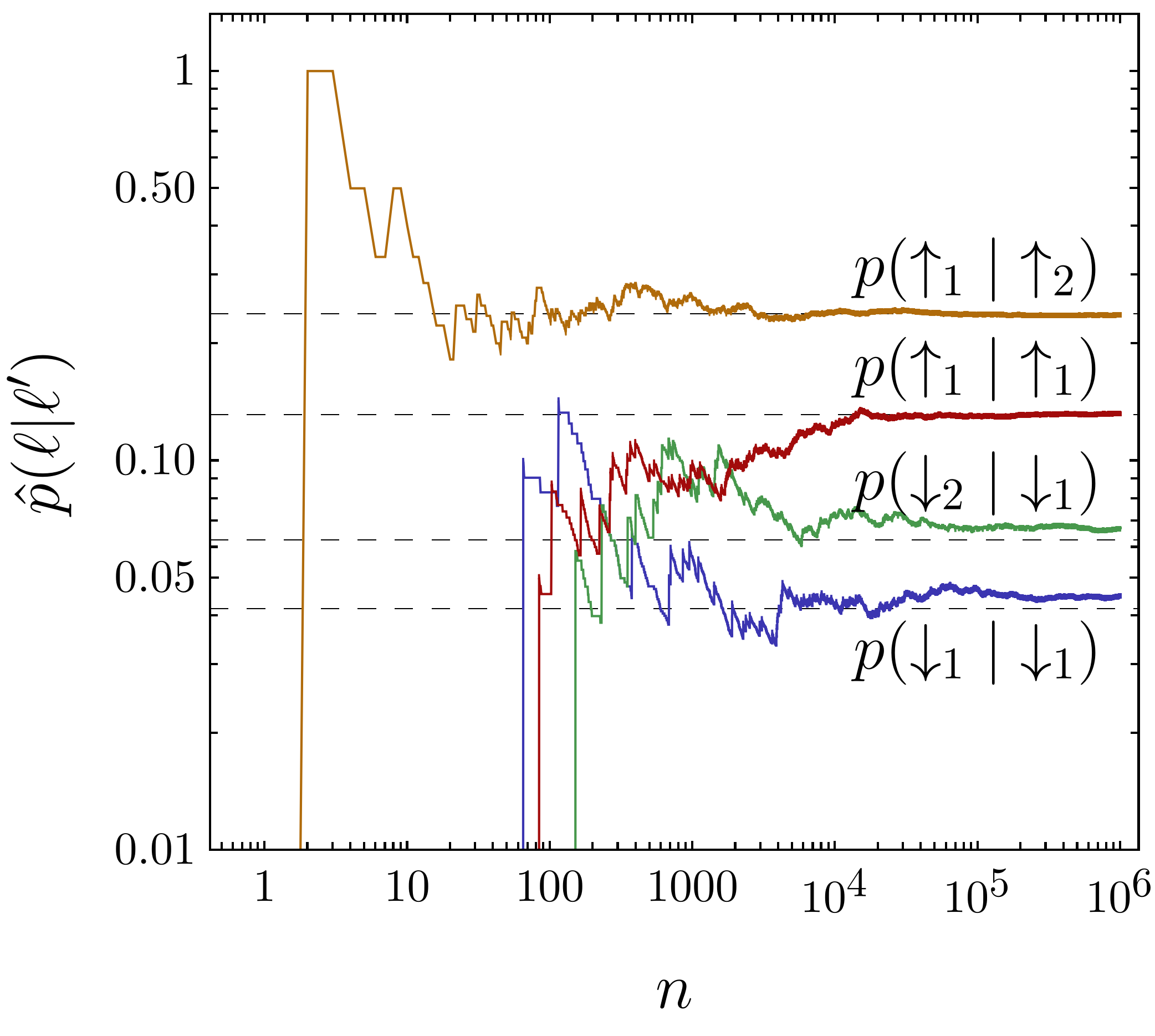}%
\includegraphics[height=.32\textheight, clip, trim = {44pt 0 0 0}]{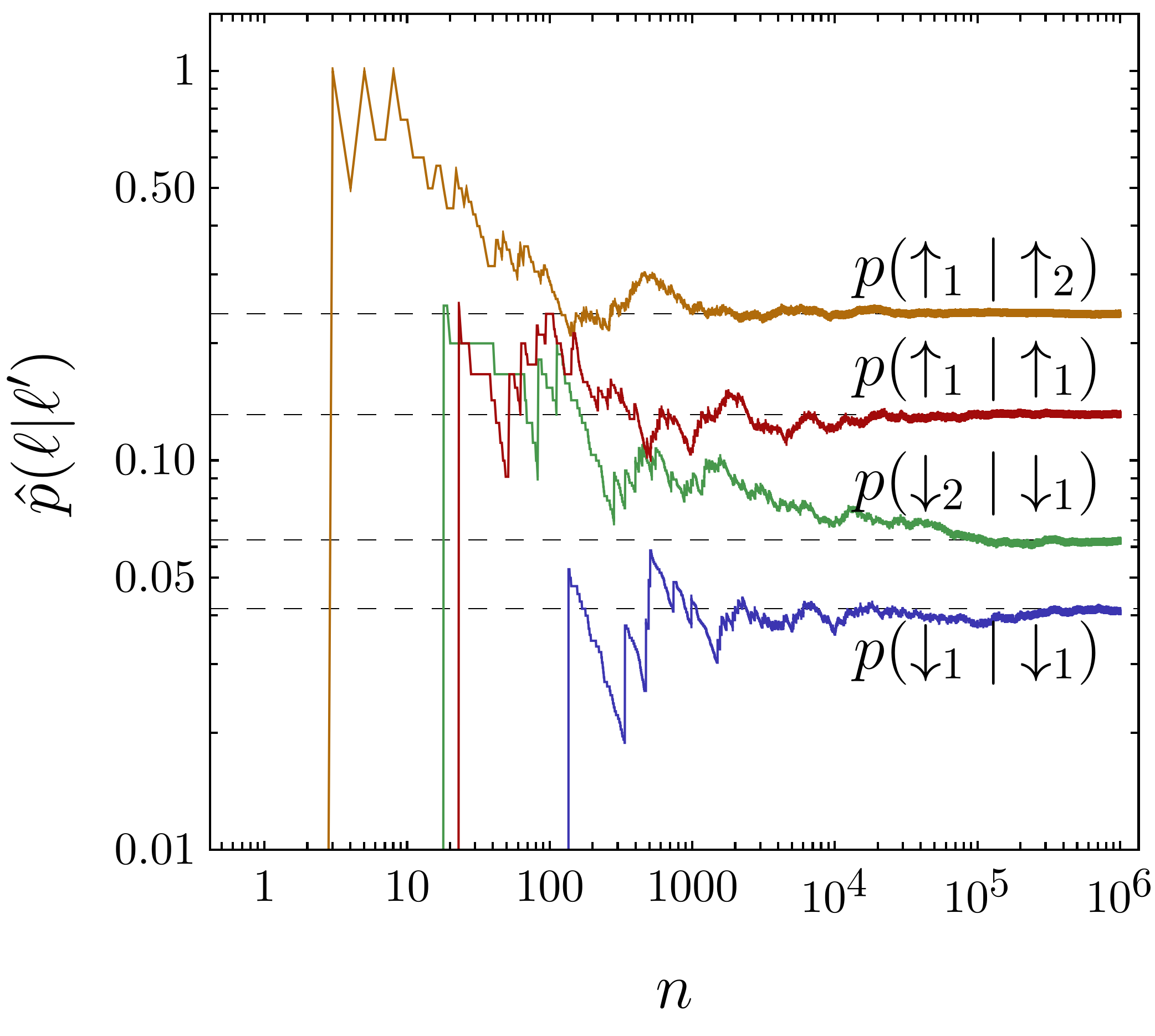}
\caption{Convergence of estimators for some trans-transition probability to the theoretical values (dashed lines) with the increase of $n$. Left: estimation using a single experiment. Right: estimation using different experiments, as in Eq.~\eqref{eq:trans-transitionestimator}.}
\label{fig:ergodic}
\end{figure}

Another important point is represented by the fact that, if a long stationary sequence of transitions $\mathcal{L}_n$ is known, it is possible to estimate trans-transition probabilities. In fact, it is possible to extract the number of times a transition $\ell$ occurs after $\ell'$, denoted $n_{\ell\ell'}(\mathcal{L}_n)$, and also the bare number of $\ell'$ occurrences $n_{\ell'}(\mathcal{L}_n)$. The trans-transition probabilities can then be estimated by
\begin{equation}
\hat{p}(\ell|\ell') = \frac{n_{\ell\ell'}(\mathcal{L}_n)}{n_{\ell'} (\mathcal{L}'_n)},
\label{eq:trans-transitionestimator}
\end{equation}
where we highlight that it is good practice to estimate them from different experiments $\mathcal{L}_n$ and $\mathcal{L}'_n$. In \fref{fig:ergodic}, we see the converge of estimations by Eq.~\eqref{eq:trans-transitionestimator} done with the same experiment (left panel) or with different experiments (right panel). In the former, autocorrelations are present and break the convergence of estimators \cite{Sokal1997}, which is more visible for $p(\down_2 \vert \down_1)$ and $p(\down_1 \vert \down_1)$. An ergodic theorem for the estimated trans-transition probabilities was not proven in this work, and was assumed to hold as suggested by \fref{fig:ergodic}, where we see the convergence of estimators to their true value.

An interesting open problem emerging from this article is the generalization of KCL to an arbitrary number of observable events. This is not a straightforward application of the known procedure in state space in terms of spanning trees and a cyclomatic number \cite{avanzini2023methods} as a mixed current is the difference of the fluxes $n_{\ell\ell'}$ at different nodes of the transition space graph.

As a final comment, other works Refs.\,\cite{marginal,effectivefluc} deal with partial currents, evaluated at clock time $t$. In the first one, the case of a single observable current is addressed, and in the second one the case of more observable currents is included. In both cases, the effective affinities are defined in terms of the stalling distribution, i.e. the stationary distribution of the original system where the observable edges are removed. Numerical evidence suggests that the effective affinities in these works correspond to the effective affinities $a_\nu$ in the cases of a single  observable current or a complete set. It is yet to be understood whether there exists a connection between the stalling states and $a_\nu$ for a set with an arbitrary number of currents.

\section{Acknowledgements}
\label{sec:acknowledgements}

The research was supported by the National Research Fund Luxembourg (project CORE ThermoComp C17/MS/11696700), by the European Research Council, project NanoThermo (ERC-2015-CoG Agreement No. 681456), and by the project INTER/FNRS/20/15074473 funded by F.R.S.-FNRS (Belgium) and FNR (Luxembourg).

\appendix

\section{}

\subsection{Relation between loop currents and total currents}

\label{app:1}

We prove that the expression for the total integrated current $c_\nu$ of kind $\nu$ evaluated for a single realization $\mathcal{L}_n$ of the process in the observable transition space in terms of loop and mixed currents is
\begin{equation}
\fl c_\nu (\mathcal{L}_n) = \xi_\nu (\mathcal{L}_n) + \partial_\nu (\ell^{(1)},\ell^{(n)}) + \frac{1}{2}\sum_{\underset{\mu < \sigma}{\mu,\sigma}} \sum_{(\ell_\mu,\ell_\sigma)} \xi_{\ell_\mu \ell_\sigma} (\mathcal{L}_n) (\delta_{\mu,\nu}j(\ell_\mu) + \delta_{\sigma,\nu}j(\ell_\sigma)).
\end{equation}
with
\begin{equation}
\eqalign{
\fl \partial_\nu (\ell^{(1)},\ell^{(n)}) = \frac{1}{2} (\delta_{\ell^{(1)}, \up_\nu} - \delta_{\ell^{(1)} , \down_\nu} + \delta_{\ell^{(n)} , \up_\nu} - \delta_{\ell^{(n)} , \down_\nu}) \\
= \frac{1}{2} \sum_{\ell_\nu} (\delta_{\ell^{(1)} , \ell_\nu} j(\ell_\nu) - \delta_{\bar{\ell}^{(n)} , \ell_\nu} j(\ell_\nu)),
}
\end{equation}
collecting the contributions at the boundaries of $\mathcal{L}_n$.

The equation above can be obtained by considering a decomposition of a sequence of observed transitions $\mathcal{L}_n$ in terms of snippets of the same kind, as explained in \sref{sec:FR}. The contribution of each snippet of kind $\nu$ to the total currents is given by Eq.\,\eqref{eq:totalcurrent} \cite{beatofacurrent}. Thus the total current of kind $\nu$ is
\begin{equation}
\fl c_\nu(\mathcal{L}_n) = \sum_{i'} c_\nu(\mathcal{L}_\nu^{(i')}) = \xi_\nu (\mathcal{L}_n) + \frac{1}{2} \sum_{i'} j(\ell_\nu^{(1_{i'})}) + \frac{1}{2} \sum_{i'} j(\ell_\nu^{(n_{i'})}) 
\label{eq:snippetsum}
\end{equation}

The second term in the RHS of Eq.\,\eqref{eq:snippetsum} contains the contributions by the mixed currents $\lbrace \xi_{\ell_\nu\ell_\mu} \rbrace$ since the number of times the first transition $j(\ell_\nu^{(1_{i'})})$ in a snippet of kind $\nu$ is $\up_\nu$ (respectively $\down_\nu$) is the number of times $\up_\nu$ ($\down_\nu$) occurs after any other transition of a different kind $\mu \neq \nu$, with an additional contribution when the first transition occurring in the full sequence $\mathcal{L}_n$ is $\up_\nu$ ($\down_\nu$). Thus
\begin{equation}
\fl \frac{1}{2} \sum_{i'} j(\ell_\nu^{(1_{i'})}) = \frac{1}{2} \sum_{\ell_\mu \neq \ell_\nu} \left( n_{\up_\nu \ell_\mu}(\mathcal{L}_n) -  n_{\down_\nu \ell_\mu}(\mathcal{L}_n) \right) + \frac{1}{2} (\delta_{\up_\nu,\ell^{(1)}} - \delta_{\down_\nu,\ell^{(1)}} ).
\label{eq:secondterm}
\end{equation}
With analogous arguments we write the third term in Eq.\,\eqref{eq:snippetsum} as
\begin{equation}
\fl \frac{1}{2} \sum_{i'} j(\ell_\nu^{(n_{i'})}) = \frac{1}{2} \sum_{\ell_\mu \neq \ell_\nu} \left( n_{\ell_\mu \up_\nu}(\mathcal{L}_n) -  n_{\ell_\mu \down_\nu}(\mathcal{L}_n) \right) + \frac{1}{2} (\delta_{\up_\nu,\ell^{(n)}} - \delta_{\down_\nu,\ell^{(n)}} ).
\label{eq:thirdterm}
\end{equation}

By plugging {Eqs.\,}\eqref{eq:secondterm} and \eqref{eq:thirdterm} into Eq.\,\eqref{eq:snippetsum}, one obtains for the total current of kind $\nu$ in a single trajectory $\mathcal{L}_n$ of length $n$
\begin{equation}
\eqalign{
\fl c_\nu (\mathcal{L}_n) = \xi_\nu (\mathcal{L}_n) + \partial_\nu (\ell^{(1)},\ell^{(n)})\\
 + \frac{1}{2}\sum_{\mu \neq \nu} \left( \sum_{\ell_\mu} n_{\up_\nu \ell_\mu} (\mathcal{L}_n) - n_{\down_\nu\ell_\mu} (\mathcal{L}_n) +  n_{\ell_\mu \up_\nu}(\mathcal{L}_n) - n_{\ell_\mu \down_\nu}(\mathcal{L}_n) \right) }
\label{eq:totcurrentn}
\end{equation}
with the boundary term $\partial_\nu(\ell^{(1)},\ell^{(n)}) = \frac{1}{2} (\delta_{\ell^{(1)},\up_\nu} - \delta_{\ell^{(1)},\down_\nu} + \delta_{\ell^{(n)},\up_\nu} - \delta_{\ell^{(n)},\down_\nu})$. Notice that the sign on each $n_{\ell_\nu\ell_\mu}(\mathcal{L}_n)$ depends on the elementary current $j(\ell_\nu)$ carried by $\ell_\nu$. 

Let us now focus on a single kind $\mu \neq \nu$. The contribution of the transitions of kind $\mu$ inside the parentheses in Eq.\,\eqref{eq:totcurrentn} is given by
\begin{equation}
n_{\up_\nu \up_\mu} - n_{\down_\nu\up_\mu} + n_{\up_\nu \down_\mu} - n_{\down_\nu\down_\mu} + n_{\up_\mu\up_\nu} - n_{\up_\mu\down_\nu} + n_{\down_\mu\up_\nu} - n_{\down_\mu\down_\nu}
\end{equation}
and we can recognize the mixed currents $\xi_{\up_\nu\up_\mu}$, $\xi_{\up_\nu\down_\mu}$, $\xi_{\down_\nu\up_\mu}$ and $\xi_{\down_\nu\down_\mu}$ with the sign fixed by the elementary current $j(\ell_\nu)$ carried by transition $\ell_\nu$. Finally, we can generalize Eq.\,\eqref{eq:snippetsum} to the case of multiple observable transitions as
\begin{equation}
\fl c_\nu(\mathcal{L}_n) = \xi_\nu(\mathcal{L}_n) + \partial_\nu (\ell^{(1)},\ell^{(n)}) + \frac{1}{2} \sum_{\underset{ \sigma > \mu}{\mu,\sigma}} \sum_{(\ell_\mu,\ell_\sigma)} \xi_{\ell_\mu\ell_\sigma} (\mathcal{L}_n) (\delta_{\mu,\nu} j(\ell_\mu) + \delta_{\sigma,\nu} j(\ell_\sigma))
\end{equation}
thus proving Eq.\,\eqref{eq:generalizedcurr}.

\subsection{Shifted affinities, shifted potential}

\label{app:plug}

We now plug Eq.\,\eqref{eq:generalizedcurr} inside the RHS of Eq.\,\eqref{eq:ratiocurrents}. In particular, we invert Eq.\,\eqref{eq:generalizedcurr} as
\begin{equation}
\fl \xi_\nu(\mathcal{L}_n) = c_\nu(\mathcal{L}_n) - \partial_\nu (\ell^{(1)},\ell^{(n)}) - \frac{1}{2} \sum_{\underset{\mu < \sigma}{\mu,\sigma}} \sum_{(\ell_\mu,\ell_\sigma)} \xi_{\ell_\mu\ell_\sigma} (\mathcal{L}_n) (\delta_{\mu,\nu} j(\ell_\mu) + \delta_{\sigma,\nu} j(\ell_\sigma)),
\end{equation}
Explicitly
\begin{eqnarray}
\fl \ln \frac{p(\mathcal{L}_n)}{p(\bar{\mathcal{L}}_n)}  = \sum_\nu a_\nu c_\nu(\mathcal{L}_n)  \label{eq:currentcontr} \\
- \frac{1}{2} \sum_\nu a_\nu \left(  \sum_{\ell_\nu} (\delta_{\ell^{(1)} , \ell_\nu} j(\ell_\nu) - \delta_{\bar{\ell}^{(n)} , \ell_\nu} j(\ell_\nu))  \right)  \label{eq:potshift}\\
- \frac{1}{2} \sum_\nu a_\nu \left(  \sum_{\underset{\sigma > \mu}{\mu,\sigma}} \sum_{(\ell_\mu,\ell_\sigma)} \xi_{\ell_\mu\ell_\sigma} (\mathcal{L}_n) (\delta_{\mu,\nu} j(\ell_\mu) + \delta_{\sigma,\nu} j(\ell_\sigma))\right) \label{eq:affshift}\\
+ \sum_{\underset{\mu > \nu}{\mu, \nu}} \sum_{(\ell_\nu, \ell_\mu)} \alpha_{\ell_\nu\ell_\mu} \xi_{\ell_\nu\ell_\mu} (\mathcal{L}_n) \label{eq:affcontr}\\
+ u(\bar{\ell}^{(n)}) - u(\ell^{(1)}) \label{eq:potcontr}.
\end{eqnarray}
with $u(\ell) = -\ln p_1(\ell)$. The first term Eq.\,\eqref{eq:currentcontr} is the contribution due to the total currents to the log-ratio Eq.\,\eqref{eq:ratiocurrents}. The second term Eq.\,\eqref{eq:potshift} can be incorporated with the last term Eq.\,\eqref{eq:potcontr}, thus defining the shifted boundary potential Eq.\,\eqref{eq:tiltedcurredpot}
\begin{equation}
\tilde{u} (\ell) = u(\ell) + \frac{1}{2} \sum_\nu \sum_{\ell_\nu} a_\nu j(\ell_\nu) \delta_{\ell,\ell_\nu} + v,
\end{equation}
with $v$ a constant which fixes the normalization of the initial probabilities. In the third term Eq.\,\eqref{eq:affshift}, we can apply the summation over kinds $\nu$, obtaining
\begin{equation}
\eqalign{
\fl - \frac{1}{2} \sum_\nu a_\nu \left(  \sum_{\underset{\sigma > \mu}{\mu,\sigma}} \sum_{(\ell_\mu,\ell_\sigma)} \xi_{\ell_\mu\ell_\sigma} (\mathcal{L}_n) (\delta_{\mu,\nu} j(\ell_\mu) + \delta_{\sigma,\nu} j(\ell_\sigma))\right) \\
 = \sum_{\underset{\sigma > \mu}{\mu,\sigma}} \sum_{(\ell_\mu,\ell_\sigma)} \left[ - \frac{1}{2}\left(  a_\mu j(\ell_\mu) + a_\sigma j(\ell_\sigma)\right) \right] \xi_{\ell_\mu\ell_\sigma} (\mathcal{L}_n) .
}
\end{equation}
By renaming the indices $\mu \to \nu$ and $\sigma \to \mu$, the expression above can be summed with Eq.\,\eqref{eq:affcontr}, thus defining the shifted mixed affinities Eq.\,\eqref{eq:tiltedcurredaff}
\begin{equation}
\tilde{\alpha}_{\ell_\nu\ell_\mu} = \alpha_{\ell_\nu\ell_\mu} - \frac{1}{2} \left( a_\nu j(\ell_\nu) + a_\mu j(\ell_\mu) \right).
\end{equation}

Finally, Eq.\,\eqref{eq:ratiocurrents} is rewritten as
\begin{equation}
\fl \ln \frac{p(\mathcal{L}_n)}{p(\bar{\mathcal{L}}_n)} = \sum_{\nu} a_\nu c_\nu (\mathcal{L}_n) + \sum_{\underset{\mu> \nu}{\mu, \nu}} \sum_{(\ell_\nu\ell_\mu)} \tilde{\alpha}_{\ell_\nu\ell_\mu} \xi_{\ell_\nu\ell_\mu} (\mathcal{L}_n) + \tilde{u} (\bar{\ell}^{(n)}) - \tilde{u} (\ell^{(1)}) ,
\end{equation}
which completes the proof.

\section{}

\label{app:genfuncurrents}

\newcommand{\cur}{c}

We express the MGF for currents Eq.\,\eqref{eq:MGFdef} in terms of the tilted trans-transition matrix Eq.\,\eqref{eq:tiltedmatrixcurr} dressed with counting fields for the case of multiple observable currents $\boldsymbol{c}$ and the mixed currents $\bxi$, thus obtaining the equivalent expression Eq.\,\eqref{eq:MGFalt}.

Let $p_n(\boldsymbol{c},\boldsymbol{\xi}, \ell)$ be the probability that the $n$-th transition is $\ell$ and that the currents take the values $\boldsymbol{c}$ and $\bxi$ after $n$ occurrences of observable transitions. Its MGF is
\begin{equation}
G_n(\boldsymbol{k},\bkappa,\ell) = \sum_{\boldsymbol{c},\bxi \in \mathcal{F}_n} e^{\boldsymbol{k}\cdot \boldsymbol{c} + \bkappa \cdot \bxi} p_n(\boldsymbol{c},\boldsymbol{\xi}, \ell) .
\end{equation}
with $\boldsymbol{k}$ denoting the vector of counting fields conjugated with the currents $\boldsymbol{c}$ and $\bkappa$ the vector of counting fields conjugated with the mixed currents $\bxi$. $\mathcal{F}_n$ here denotes the set of possible values the currents can simultaneously take at time $n$. Notice that the first transition $\ell^{(1)}$ contributes $j(\ell^{(1)})$ to the current of its kind, therefore
\begin{equation}
G_1(\boldsymbol{k}, \bkappa, \ell_\nu) =  e^{k_\nu j(\ell_\nu)} p_1(\ell_\nu) = [E(\boldsymbol{k})]_{\ell_\nu,\ell_\nu} p_1(\ell_\nu). 
\label{eq:initgenfun}
\end{equation}
with $E$ the matrix defined by Eq.\,\eqref{eq:matrixE}, if the first transition is of kind $\nu$.

After $n$ transitions the MGF $G_n(\boldsymbol{k},\bkappa)$ is then defined as
\begin{equation}
G_n(\boldsymbol{k},\bkappa) =  \sum_\ell G_n(\boldsymbol{k},\bkappa,\ell) =  \sum_\nu \sum_{\ell_\nu} G_n(\boldsymbol{k},\bkappa,\ell_\nu),
\end{equation}
where we highlighted the kind of the last occurring transition $\ell^{(n)}$ at time $n$.

To find an expression for it, first we look for an evolution equation for $G_n(\boldsymbol{k},\bkappa,\ell_\nu)$. From $\boldsymbol{p}_{n+1} = P \boldsymbol{p}_n$ we have
\begin{eqnarray}
\fl G_{n+1}(\boldsymbol{k},\bkappa,\ell_\nu) = \sum_{\boldsymbol{c},\bxi \in \mathcal{F}_{n+1}} e^{\boldsymbol{k}\cdot \boldsymbol{c} + \bkappa \cdot \bxi} p_{n+1}(\boldsymbol{c},\boldsymbol{\xi}, \ell_\nu) \label{eq:evolutionMGF}\\
= \sum_\mu \sum_{\ell_\mu'} p(\ell_\nu|\ell_\mu')   \sum_{\boldsymbol{c},\bxi   \in \mathcal{F}_{n+1}}  e^{\boldsymbol{k}\cdot \boldsymbol{c} + \bkappa \cdot \bxi} \times \nonumber \\ 
\times p_{n}(\lbrace c_\sigma - j(\ell_\nu)\delta_{\sigma,\nu} \rbrace, \lbrace \xi_{\ell_\sigma\ell_\rho} - (1-\delta_{\nu,\mu}) (\delta_{\ell_\sigma, \ell_\nu} \delta_{\ell_\rho, \ell_\mu'} - \delta_{\ell_\sigma, \bar{\ell}_\mu'} \delta_{\ell_\rho, \bar{\ell}_\nu} ) \rbrace , \ell_\mu'), \nonumber
\end{eqnarray}
where notice that the last trans-transition $\ell_\mu' \to \ell_\nu$ increases the current $c_\nu$ by $j(\ell_\nu)$ and the mixed current $\xi_{\ell_\nu\ell_\mu'}$ ($\xi_{\bar{\ell}_\mu' \bar{\ell}_\nu}$) by $1$ ($-1$) when $\mu \neq \nu$. 

Given that the last transition is $\ell_\nu$, the set of possible values that the current $c_\nu$ can assume at time $n+1$ includes all values it could have taken before, $\mathcal{F}_n^{c_\nu}$, and also the new elements $\pm (n+1)$; conveniently, we can also denote it by
\begin{equation}
\mathcal{F}_{n+1}^{c_\nu} = \{\mathcal{F}_n^{c_\nu} - 1\} \cup \{n,n+1\} = \{\mathcal{F}_n^{c_\nu} + 1\} \cup \{-n,-n-1\} ,
\label{eq:filtrations}
\end{equation}
and the possible outcomes of all the other currents of kind $\mu\neq \nu$ are unchanged with respect the previous time $n$. The same happens for the mixed currents since the $n+1$-th and $n$-th transitions are of the same kind.

We now consider $\ell_\nu = \down_\nu$, and distinguish between the two cases where $\mu = \nu$ and $\mu \neq \nu$ in Eq.\,\eqref{eq:evolutionMGF}. In the first case we have that the values $c_\nu = n$ and $c_\nu = n+1$ do not contribute because the probability that $c_\nu = n+1$ or $c_\nu = n+2$ at time $n$ is zero. By using the first equality in Eq.\,\eqref{eq:filtrations} and denoting with $\boldsymbol{c}'$ the vector of currents that are not of kind $\nu$, we express the terms in the sum Eq.\,\eqref{eq:evolutionMGF} with $\mu = \nu$ as
\begin{equation}
\sum_{\ell_{\mu=\nu}'} p(\down_\nu|\ell_\mu') \sum_{c_\nu + 1 \in \mathcal{F}_n^{c_\nu}} \sum_{\boldsymbol{c}',\bxi \in \mathcal{F}_n'} e^{\boldsymbol{k}\cdot \boldsymbol{c} + \bkappa \cdot \bxi} p_{n}(c_\nu + 1, \boldsymbol{c}', \bxi, \ell_\mu')  ,
\end{equation}
with $\mathcal{F}_n'$ denoting the set of possible values the joint set of the currents $\boldsymbol{c}'$ and $\bxi$ can take at time $n$.
By shifting the current $c_\nu$ we obtain
\begin{equation}
\fl \sum_{\ell_{\mu=\nu}'}  p(\down_\nu|\ell_\mu') e^{-k_\nu} \sum_{\boldsymbol{c},\bxi \in \mathcal{F}_n} e^{\boldsymbol{k}\cdot \boldsymbol{c} + \bkappa \cdot \bxi} p_{n}(\boldsymbol{c},\bxi , \ell_\mu') = \sum_{\ell_{\mu=\nu}'} [P(\boldsymbol{k},\bkappa)]_{\down_\nu,\ell_\mu'} G_n(\boldsymbol{k}, \bkappa, \ell_\mu'),
\end{equation}
where we identified the definition of $G_n(\boldsymbol{k},\bkappa ,\ell_\mu')$ and the diagonal blocks of the tilted matrix Eq.\,\eqref{eq:tiltedmatrixcurr}. 

In the remaining case of $\mu\neq \nu$ the mixed current $\xi_{\down_\nu\ell_\mu'}$ ($\xi_{\bar{\ell}_\mu' \bar{\ell}_\nu}$) is also increased (decreased), while all other mixed currents are kept untouched. With similar arguments, we then have again that
\begin{equation}
\eqalign{
\fl  \sum_{\ell_{\mu\neq\nu}'} p(\down_\nu|\ell_\mu') \sum_{\boldsymbol{c},\bxi \in \mathcal{F}_n} e^{\boldsymbol{k}\cdot \boldsymbol{c} + \bkappa \cdot \bxi} p_{n}(\boldsymbol{c},\bxi , \ell_\mu') \\
= \sum_{\ell_{\mu\neq\nu}'}  p(\down_\nu|\ell_\mu') e^{-k_\nu} e^{\kappa_{\ell_\nu\ell_{\mu'}}}G_n(\boldsymbol{k}, \bkappa, \ell_\mu') \\
= \sum_{\ell_{\mu\neq\nu}'}  [P(\boldsymbol{k},\bkappa)]_{\down_\nu,\ell_\mu'} G_n(\boldsymbol{k}, \bkappa, \ell_\mu') ,
}
\end{equation}
thus
\begin{equation}
G_{n+1}(\boldsymbol{k},\bkappa,\down_\nu)  = \sum_\mu\sum_{\ell_\mu'}  [P(\boldsymbol{k},\bkappa)]_{\down_\nu,\ell_\mu'} G_n(\boldsymbol{k}, \bkappa, \ell_\mu') ,
\end{equation}
where the sum runs over all kinds $\mu$. Similarly, by considering $\ell_\nu = \up_\nu$ and by employing the second identity in Eq.\,\eqref{eq:filtrations}, we find that
\begin{equation}
G_{n+1}(\boldsymbol{k},\bkappa,\up_\nu)  = \sum_\mu\sum_{\ell_\mu'}  [P(\boldsymbol{k},\bkappa)]_{\up_\nu,\ell_\mu'} G_n(\boldsymbol{k}, \bkappa, \ell_\mu') .
\end{equation}

With $\boldsymbol{G}_{n}(\boldsymbol{k},\bkappa) = (G_{n}(\boldsymbol{k},\bkappa, \up_1), G_{n}(\boldsymbol{k},\bkappa, \down_1),\cdots, G_{n}(\boldsymbol{k},\bkappa, \up_M), G_{n}(\boldsymbol{k},\bkappa, \down_M) )$, for $M$ observable kinds, we can write
\begin{equation}
\boldsymbol{G}_{n+1}(\boldsymbol{k},\bkappa) = P(\boldsymbol{k},\bkappa) \boldsymbol{G}_{n}(\boldsymbol{k},\bkappa) ,
\end{equation}
and we remind that $P(\boldsymbol{k},\bkappa)$ denotes the tilted trans-transition matrix Eq.\,\eqref{eq:tiltedmatrixcurr}. Therefore we propagate the expression above as
\begin{equation}
\fl G_n(\boldsymbol{k},\bkappa) =  \sum_\ell G_n(\boldsymbol{k},\bkappa,\ell) = \boldsymbol{1} \cdot \boldsymbol{G}_{n}(\boldsymbol{k},\bkappa) = \boldsymbol{1} \cdot P(\boldsymbol{k},\bkappa)^{n-1} \boldsymbol{G}_1(\boldsymbol{k}, \bkappa) ,
\label{eq:momn}
\end{equation}
where $\boldsymbol{1}$ is the unitary vector with $2M$ components, and with the elements of the vector $\boldsymbol{G}_1(\boldsymbol{k})$ given by Eq.\,\eqref{eq:initgenfun} in terms of initial probabilities in transition space. Finally,
\begin{equation}
    G_n(\boldsymbol{k},\bkappa) = \boldsymbol{1} \cdot P(\boldsymbol{k},\bkappa)^{n-1} E(\boldsymbol{k}) \boldsymbol{p}_1,
\end{equation}
which concludes the proof.

\section{}

\label{app:0}

We prove the expression
\begin{equation}
E(\boldsymbol{k}) J = J E(- \boldsymbol{k}).
\label{eq:commutation}
\end{equation}
involving the matrix $E(\boldsymbol{k})$ defined by Eq.\,\eqref{eq:matrixE} and the swapping matrix $J$. The identity is immediately proven by observing that application of $J$ on the left and right of a matrix swaps pairs of rows and columns. Then
\begin{equation}
\left[ J E(\boldsymbol{k}) J \right]_{\ell_\nu\ell_\mu'} = \exp\left( - k_\nu j(\ell_\nu) \right) \delta_{\ell_\nu,\ell_\mu'} = E (-\boldsymbol{k})_{\ell_\nu\ell_\mu'}.
\end{equation}

\section{}

\label{app:3}

\subsection{Symmetry for the tilted matrix}

\label{app:3symmetry}

We seek for a diagonal matrix $D_{\ell_\nu\ell_\mu'} = d(\ell_\nu) \delta_{\ell_\nu,\ell_\mu'}$ such that Eq.\,\eqref{eq:symcurr} is satisfied. By considering the vectors $\boldsymbol{c}$ and $\bxi$ containing the total currents $c_\nu$ and $\xi_{\ell_\nu\ell_\mu}$ respectively, we consider the tilting Eq.\,\eqref{eq:tiltedmatrixcurr} for the tilted matrix $P(\boldsymbol{k},\bkappa)$. For $\nu = \mu$ we get the conditions
\begin{eqnarray}
\frac{p(\ell_\nu|\ell_\nu)}{p(\bar{\ell}_\nu|\bar{\ell}_\nu)} = \exp(a_\nu j(\ell_\nu)) \qquad & \ell_\nu = \ell_\nu'\\
\label{eq:condition1}
\frac{d(\ell_\nu)}{d(\bar{\ell}_\nu)} = \exp(- a_\nu j(\ell_\nu) ) \qquad & \ell_\nu = \bar{\ell}_\nu'.
\label{eq:condition2}
\end{eqnarray}
For $\nu \neq \mu$ we have
\begin{equation}
\frac{d(\ell_\mu)}{d(\ell_\nu)} = \frac{p(\ell_\nu|\ell_\mu)}{p(\bar{\ell}_\mu|\bar{\ell}_\nu)}\exp\left(-a_\mu j(\ell_\mu) - \tilde{\alpha}_{\ell_\nu\ell_\mu}\right).
\end{equation}
By using Eq.\,\eqref{eq:condition2} and the definition Eq.\,\eqref{eq:tiltedcurredaff} we get to
\begin{equation}
\frac{d(\ell_\nu)}{d(\ell_\mu)} = \exp\frac{1}{2}(a_\mu j(\ell_\mu) - a_\nu j(\ell_\nu) ).
\label{eq:condition3}
\end{equation}
We look for $d(\ell)$ which satisfy conditions Eq.\,\eqref{eq:condition2} and Eq.\,\eqref{eq:condition3} simultaneously. From condition Eq.\,\eqref{eq:condition2} one obtains that $d(\ell_\nu) = b_\nu p(\bar{\ell}_\nu|\bar{\ell}_\nu)$, with $b_\nu$ a proportionality constant depending on the type $\nu$, for all $\nu$. Condition Eq.\,\eqref{eq:condition3} can now be expressed as
\begin{equation}
\frac{d(\ell_\nu)}{d(\ell_\mu)} = \frac{p(\bar{\ell}_\nu|\bar{\ell}_\nu)}{p(\bar{\ell}_\mu|\bar{\ell}_\mu)}\frac{b_\nu}{b_\mu} = \exp\left( \frac{1}{2}(a_\mu j(\ell_\mu) - a_\nu j(\ell_\nu) \right)
\end{equation}
providing
\begin{equation}
\frac{b_\nu}{b_\mu} = \left(\frac{p(\up_\mu|\up_\mu)) p(\down_\mu|\down_\mu)}{p(\up_\nu|\up_\nu)) p(\down_\nu|\down_\nu)}\right)^{\frac{1}{2}}.
\label{eq:kratio}
\end{equation}
Since Eq.\,\eqref{eq:kratio} must hold for all pairs  $(\nu,\mu)$ we get
\begin{equation}
b_\nu \propto \prod_{\mu \neq \nu} \left[ p(\up_\mu|\up_\mu) p(\down_\mu|\down_\mu) \right]^{\frac{1}{2}}
\end{equation}
and finally
\begin{equation}
\eqalign{
d(\ell_\nu) & \propto p(\bar{\ell}_\nu|\bar{\ell}_\nu) \prod_{\mu \neq \nu} \left[ p(\up_\mu|\up_\mu) p(\down_\mu|\down_\mu) \right]^{\frac{1}{2}}\\
& =  \exp \left(-\frac{1}{2} a_\nu j(\ell_\nu) \right) \prod_{\mu}\left[ p(\up_\mu|\up_\mu) p(\down_\mu|\down_\mu) \right]^{\frac{1}{2}} ,
}
\label{eq:diagmatrixelem}
\end{equation}
where in the second row we multiplied and divided the previous expression by $p(\ell_\nu|\ell_\mu)$. It is immediate to verify that Eq.\,\eqref{eq:diagmatrixelem} satisfies both conditions Eq.\,\eqref{eq:condition2} and Eq.\,\eqref{eq:condition3}. Since each element of $D$ is positive, real, and not dependent on $\boldsymbol{k}$ or $\boldsymbol{\kappa}$ we conclude that such a choice satisfies Eq.\,\eqref{eq:symcurr}.

\subsection{Symmetry for the tilted matrix (complete sets)}

\label{app:3complete}

For complete sets we want the condition Eq.\,\eqref{eq:symcurr} to hold for the counting fields $\boldsymbol{k}$ associated with the observable currents $c_\nu$ only. By writing Eq.\,\eqref{eq:symcurr} explicitly we have that
\begin{eqnarray}
\left[ E(-\boldsymbol{k}) P(\boldsymbol{k},\bkappa) E(\boldsymbol{k}) \right]_{\ell_\nu,\ell_\mu'} = p(\ell_\nu |\ell_\mu') \exp (k_\mu j(\ell_\mu '))	
\label{eq:LHSapp} \\
\left[ D^{-1} P(-\boldsymbol{a} - \boldsymbol{k},-\tilde{\balpha} - \bkappa^\perp D \right]_{\ell_\nu,\ell_\mu'} = \frac{p(\bar{\ell}_\mu '|\bar{\ell}_\nu) d(\ell_\mu')}{d(\ell_\nu)} \exp ((a_\mu + k_\mu) j(\ell_\mu '))
\label{eq:RHSapp}
\end{eqnarray}
thus finding
\begin{equation}
p(\ell_\nu |\ell_\mu')=  \frac{p(\bar{\ell}_\mu '|\bar{\ell}_\nu) d(\ell_\mu')}{d(\ell_\nu)} \exp (a_\mu j(\ell_\mu ' )) .
\label{eq:condition1app}
\end{equation}
In the case where $\ell_\nu	 = \ell_\mu'$ we recover the definition for the affinities $a_\nu$:
\begin{equation}
a_\nu = \frac{p(\up_\nu|\up_\nu)}{p(\down_\nu|\down_\nu)}.
\end{equation}
When $\ell_\nu = \bar{\ell}_\mu'$ we have instead
\begin{equation}
\frac{d(\ell_\nu)}{d(\bar{\ell}_\nu)} = \exp( - a_\nu j(\ell_\nu) )
\label{eq:condition2app}
\end{equation}
which is Eq.\,\eqref{eq:prop2}. Finally, for $\nu\neq \mu$, by plugging Eq.\,\eqref{eq:condition2app} inside Eq.\,\eqref{eq:condition1app} and swapping $\ell_\nu \leftrightarrow \bar{\ell}_\nu$ one finds Eq.\,\eqref{eq:prop1} as
\begin{equation}
\frac{d(\ell_\nu)}{d(\ell_\mu)} = \frac{p(\ell_\mu|\bar{\ell}_\nu)}{p(\ell_\nu|\bar{\ell}_\mu)} = \exp( - \alpha_{\ell_\nu \bar{\ell}_\mu } ) .
\label{eq:condition3app}
\end{equation}
As discussed in the main text, Eq.\,\eqref{eq:condition3app} must hold for all $\nu$ and $\mu$. This is verified for complete sets since the graph where the observable edges are removed is at equilibrium, containing no cycles. Consequently to Eq.\,\eqref{eq:equilibrium} which states that the intertype process is at equilibrium in the case of complete currents, we can write the shifted mixed affinities as
\begin{equation}
\tilde{\alpha}_{\nu\mu} = \upsilon_\nu - \upsilon_\mu
\end{equation}
for potentials $\upsilon_\nu$ and $\upsilon_\mu$ associated with kinds $\nu$ and $\mu$ respectively. Notice that the expression for $\tilde{\alpha}_{\nu\mu}$ is invariant under a constant shift $\upsilon_\nu \to \upsilon_\nu + v$, with $v$ a constant, for all $\nu$. Now we can proceed as in the previous section, since we can write each element $d(\ell)$ as a function of $\ell$ and its type, obtaining that the diagonal matrix $D$ satisfying conditions Eq.\,\eqref{eq:condition1app}, Eq.\,\eqref{eq:condition2app} and Eq.\,\eqref{eq:condition3app} has elements
\begin{equation}
d(\ell_\nu) \propto \exp\left(-\frac{1}{2}a_\nu j(\ell_\nu) -  \upsilon_\nu\right)  \prod_{\mu} \left[ p(\up_\mu|\up_\mu) p(\down_\mu|\down_\mu) \right]^{\frac{1}{2}}
\label{eq:matrixcomplete}
\end{equation}

\subsection{Preferred initial distribution}

\label{app:3preferred}

By using the result Eq.\,\eqref{eq:diagmatrixelem} and the fact that that the preferred initial distribution is found according to  Eq.\,\eqref{eq:preferreddef} we find after some manipulations
\begin{equation}
p_{1,{\rm nc}}^*(\ell_\nu) \propto \frac{1}{d(\ell_\nu)} = \frac{\exp\left(\frac{1}{2} a_\nu j(\ell_\nu)\right)}{\sqrt{\prod_\mu p(\ell_\mu|\ell_\mu)p(\bar{\ell}_\mu |\bar{\ell}_\mu)}}.
\end{equation}
By normalizing
\begin{equation}
p_{1,{\rm nc}}^*(\ell_\nu) = \frac{\exp\left(\frac{1}{2} a_\nu j(\ell_\nu)\right)}{2\sum_{\mu} \cosh \left( \frac{a_\mu}{2} \right)}.
\end{equation}
which is the preferred initial distribution in the case of a non-complete set of currents. For complete sets, we use the solution Eq.\,\eqref{eq:matrixcomplete} thus finding after similar passages
\begin{equation}
p_{1, {\rm c}}^*(\ell_\nu) = \frac{\exp(\frac{1}{2}a_\nu j(\ell_\nu) + \upsilon_\nu)}{2\sum_\mu \exp(\upsilon_\mu) \cosh\left( \frac{a_\mu}{2} \right)}
\end{equation}

\section{}

\label{app:hiddenpathcomplete}

Here we prove that, for a complete set of observable transitions, the trans-transition probabilities can be factorized as 
\begin{equation}
p(\ell|\ell') = \sigma(\ell|\ell') \pi(\Gamma(\mathtt{t}(\ell') \to \mathtt{s}(\ell))) \pi(\ell)
\label{eq:factorization}
\end{equation}
with $\pi(\ell):= \pi(\mathtt{t}(\ell)|\mathtt{s}(\ell))$, $\Gamma(\mathtt{t}(\ell') \to \mathtt{s}(\ell))$ the shortest path connecting $\mathtt{t}(\ell')$ to $\mathtt{s}(\ell)$, and $\sigma(\ell|\ell')$ accounting for the excursions from the main path $\Gamma$. Moreover, given the time-reversed occurrence $\bar{\ell} \to \bar{\ell}'$, the respective probability $p(\bar{\ell}'|\bar{\ell})$ is written in terms of $\sigma(\bar{\ell}'|\bar{\ell})$ which is symmetric with respect to our notion of time-reversal which inverts both the order and the direction of observable transitions. Therefore we show that
\begin{equation}
\sigma(\ell|\ell') = \sigma(\bar{\ell}'|\bar{\ell}) ,
\label{eq:symmetryexcursions}
\end{equation}
for all $\ell,\ell' \in \bigcup_{\nu}\lbrace \up_\nu,\down_\nu \rbrace$, $\nu = 1,\dots,M$.

\def\length{1.5}

\begin{figure}
\centering
\includegraphics[scale=1]{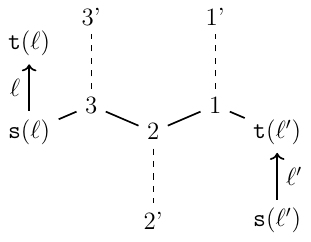}
\caption{When the set of observable currents is complete, a sequence of visible transition $\ell \to \ell'$ is connected by the unique shortest path $\Gamma(\mathtt{t}(\ell') \to \mathtt{s}(\ell))$, i.e. with the least number of transitions, which is indicated by thick lines. Fixed $\ell$ and $\ell'$, the dashed lines represent the futile branches of the network, after cancelation of all other observable edges.}
\label{fig:branches}
\end{figure}

To begin, we consider an event where $\ell'$ is followed by $\ell$. The trans-transition probability, in virtue of Eq.\,\eqref{eq:transtransition} contains the contributions of all the hidden paths starting from $\mathtt{t}(\ell')$ and ending in $\mathtt{s}(\ell)$ \cite{beatofacurrent}. As in all these contributions, the direct path $\Gamma(\mathtt{t}(\ell') \to \mathtt{s}(\ell))$ has to be performed at least once, and since it is unique it can be factorized, thus proving that trans-transition probabilities can be written in the form Eq.\,\eqref{eq:factorization}.

For the second property Eq.\,\eqref{eq:symmetryexcursions}, we reference to \fref{fig:branches}, and consider the trajectory
\begin{equation}
\fl \Xi(\ell|\ell') = \mathtt{t}(\ell') \to 1 \to \mathtt{t}(\ell')  \to 1 \to 1' \to 1 \to 2 \to 3 \to 3' \to 3 \to \mathtt{s}(\ell) \to \mathtt{t}(\ell) ,
\end{equation}
which contributes to the trans-transition probability $p(\ell|\ell')$. In fact, indicating with $\Gamma (\mathtt{t}(\ell') \to \mathtt{s}(\ell)) = \mathtt{t}(\ell') \to 1 \to 2 \to 3 \to \mathtt{s}(\ell)$, the probability of this single contribution is
\begin{equation}
\eqalign{
\fl p(\Xi(\ell|\ell')) = \pi(\ell) p(\Gamma(\mathtt{t}(\ell') \to \mathtt{s}(\ell))) \left[ \pi(1|\mathtt{t}(\ell')) \pi(\mathtt{t}(\ell')) \pi(1|1') \pi(1'|1) \pi(3|3') \pi(3'|3) \right]\\
\pi(\ell) p(\Gamma(\mathtt{t}(\ell') \to \mathtt{s}(\ell))) \sigma(\Xi(\ell|\ell')) ,
}
\end{equation}
with $\sigma(\Xi(\ell|\ell'))$ gathering the contributions due to the excursions from the main path $\Gamma$. The trans-transition probability $p(\ell|\ell')$ can then be obtained by summation over all $\Xi$, and since the first two terms can be factorized, it only affects the excursions $\sigma(\Xi)$. Therefore
\begin{equation}
\sigma (\ell|\ell') = \sum_{\Xi} \sigma (\Xi (\ell|\ell')) .
\end{equation}
Let us now consider the path 
\begin{equation}
\fl \tilde{\Xi}(\bar{\ell}'|\bar{\ell}) = \mathtt{s}(\ell) \to 3\to 3' \to 3\to 2 \to 1\to 1' \to 1 \to \mathtt{t}(\ell') \to 1 \to \mathtt{t}(\ell') \to \mathtt{s} (\ell') ,
\end{equation}
which contributes to the time-reversed trans-transition probability $p(\bar{\ell}'|\bar{\ell})$. Its probability is
\begin{equation}
\eqalign{
\fl p(\tilde{\Xi}(\bar{\ell}'|\bar{\ell})) = p(\bar{\ell}') p(\Gamma(\mathtt{s}(\ell) \to \mathtt{t}(\ell'))) \left[ \pi(3|3')\pi(3'|3) \pi(1'|1)\pi(1|1') \pi(\mathtt{t}(\ell')|1) \pi(1|\mathtt{t}(\ell')) \right] \\
= p(\bar{\ell}') p(\Gamma(\mathtt{s}(\ell) \to \mathtt{t}(\ell'))) \sigma(\tilde{\Xi}(\bar{\ell}'|\bar{\ell})) ,
}
\end{equation}
where we notice that $\sigma(\tilde{\Xi}(\bar{\ell}'|\bar{\ell})) = \sigma(\Xi(\ell|\ell'))$. As there is a one-to one correspondence between the excursion probabilities in the forward path and the time-reversed path, then we conclude that
\begin{equation}
\sigma (\ell|\ell') = \sum_{\Xi} \sigma (\Xi (\ell|\ell')) = \sum_{\tilde{\Xi}} \sigma(\tilde{\Xi}(\bar{\ell}'|\bar{\ell})) = \sigma (\bar{\ell}'|\bar{\ell}) ,
\end{equation}
which proves Eq.\,\eqref{eq:symmetryexcursions}.

\section{}

\label{app:6}

\label{app:proofcoefficients}

\begin{figure}
\centering
\includegraphics[scale=1]{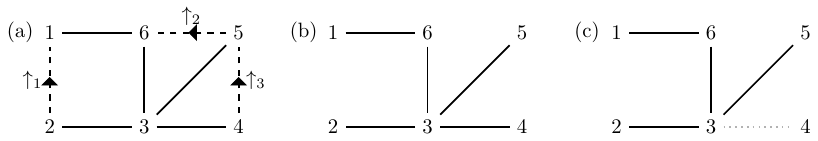}

\caption{(a) A process with 3 observable transitions over oriented edges $\uparrow_1$, $\uparrow_2$ and $\uparrow_3$ denoted with dashed lines. (b) The set of transition is complete since the graph obtained by removal of observable transitions is a tree containing no cycles. (c) The reduced spanning tree $\mathscr{T}_{12}$ is obtained by eliminating all branches that do not belong to cycles of kind $1$ or $2$; in this case, only the edge 3-4. When kinds $1$ and $3$ are considered, the reduced spanning tree is equivalent to the spanning tree (b).}
\label{fig:reducedtrees}
\end{figure}

Here, we prove Eq.\,\eqref{eq:coeffs} holds in the case of complete sets of currents since
\begin{equation}\label{eq:omeganumu}
	\omega_{\nu\mu} = \frac{ p(\ell_\nu \vert \ell_\mu) p(\bar{\ell}_\mu \vert \bar{\ell}_\mu)}{ p(\bar{\ell}_\mu \vert \bar{\ell}_\nu) p(\ell_\nu \vert \ell_\nu)  }
\end{equation}
assumes the same value regardless of the directions of $\ell_\mu$ and $\ell_\nu$.

We consider a pair of kinds $(\nu,\mu)$, whose cycles $\mathcal{C}_\nu$ and $\mathcal{C}_\mu$ can be obtained by introducing the respective edge to the spanning tree and removing all branches not belonging to the cycle \cite{schnakenberg}. The reduced spanning tree $\mathscr{T}_{\nu\mu}$ is obtained by combining the four shortest paths connecting sources and targets of both transitions and making the edges undirected. This reduced tree can be interpreted as the full tree after removal of branches that do not belong to the cycles formed by these two kinds nor to the connecting path between them (see \fref{fig:reducedtrees}), and is an important tool to assess trans-transition probabilities involving these kinds.

When both cycles have no edge in common, we call bridge $\mathcal{B}$ the set of edges that are left in $\mathscr{T}_{\nu\mu}$ after the removal of $\mathcal{C}_\nu$ and $\mathcal{C}_\mu$. This bridge is the unique set of edges connecting both cycles and will have to be visited if transitions of different kinds occur in sequence. Notice that the bridge might be supported by a single state, and for simplicity we still call it a bridge. We assume that edges in $\mathcal{B}$ are oriented in the direction of $\mathcal{C}_\nu$ to $\mathcal{C}_\mu$, with no loss of generality.

The cycles in question can be decomposed as
\begin{equation}
\mathcal{C}_\nu = \uparrow_\nu + \mathcal{C}^{\rm{out}}_\nu + \mathcal{C}^{\rm{in}}_\nu  \qquad \qquad \mathcal{C}_\mu = \uparrow_\mu + \mathcal{C}^{\rm{out}}_\mu + \mathcal{C}^{\rm{in}}_\mu,
\label{eq:decompositionbridge}
\end{equation}
where $ \mathcal{C}^{\rm{out}}_\nu$ is the path $\mathtt{t}(\up_\nu) \to \mathtt{s}(\mathcal{B})$ and $\mathcal{C}^{\rm{in}}_\nu : \mathtt{s}(\mathcal{B})\to \mathtt{s}(\up_\nu)$, while $ \mathcal{C}^{\rm{out}}_\mu : \mathtt{t}(\up_\mu) \to \mathtt{t}(\mathcal{B})$ and $\mathcal{C}^{\rm{in}}_\mu: \mathtt{t}(\mathcal{B})\to \mathtt{s}(\up_\mu)$.
Notice that if, for example, $\mathtt{t}(\up_\nu)$ belongs to the bridge, then $\mathcal{C}^{\rm{out}}_\nu$ is empty.

Recalling Eq.\,\eqref{eq:factorization}, we can write trans-transitions probabilities as
\begin{equation}
	p(\up_\nu|\up_\mu) = \sigma(\up_\nu|\up_\mu) \pi(\up_\nu) \pi(\mathcal{C}^{\rm{in}}_\nu) \pi(\bar{\mathcal{B}}) \pi(\mathcal{C}^{\rm{out}}_\mu),
\end{equation}
where the bar represents the reversal of edges in a sub-circuit. If the same is done for all other combinations of sequences involving these two kinds, it can be observed that Eq.\,\eqref{eq:omeganumu} always satisfies
\begin{equation}\label{eq:omeganumu_universal}
	\omega_{\nu\mu} = \frac{\sigma(\up_\mu \vert \up_\mu) }{ \sigma(\up_\nu \vert \up_\nu) }
	\frac{ \pi(\mathcal{C}^{\rm{out}}_\mu) \pi(\bar{\mathcal{C}}^{\rm{in}}_\mu) }{ \pi(\mathcal{C}^{\rm{out}}_\nu) \pi(\bar{\mathcal{C}}^{\rm{in}}_\nu) } \frac{\pi (\mathcal{B})}{ \pi(\bar{\mathcal{B}}) },
\end{equation}
where Eq.\,\eqref{eq:symmetryexcursions} has been used. The direction of the considered transitions are not relevant for this expressions, only their kinds. When the bridge is supported by a single state, $\pi(\mathcal{B}) = \pi(\bar{\mathcal{B}}) = 1$.

In the case where the bridge is not present, there is at least one shared edge between cycles $\mathcal{C}_\nu$ and $\mathcal{C}_\mu$. In this case, we can still decompose each cycle in a way such that each trans-transition probability can be written in terms of these sub-circuits. The trick is to consider the set of shared edges as a bridge $\mathcal{B}$, whose endpoints are now states belonging to both cycles.

Following the same arguments as above, the cycles can be decomposed as
\begin{equation}
\mathcal{C}_\nu = \up_\nu + \mathcal{C}^{\rm{out}}_\nu + \mathcal{B} + \mathcal{C}^{\rm{in}}_\nu  \qquad \qquad \mathcal{C}_\mu = \up_\mu + \mathcal{C}^{\rm{out}}_\mu + \mathcal{B} + \mathcal{C}^{\rm{in}}_\mu
\label{eq:decompositionshared}
\end{equation}
when the shared edges (bridge) have the same orientation in both cycles, otherwise
\begin{equation}
\mathcal{C}_\nu = \up_\nu + \mathcal{C}^{\rm{out}}_\nu + \mathcal{B} + \mathcal{C}^{\rm{in}}_\nu  \qquad \qquad \mathcal{C}_\mu = \up_\mu + \mathcal{C}^{\rm{out}}_\mu + \bar{\mathcal{B}} + \mathcal{C}^{\rm{in}}_\mu.
\label{eq:decompositionshared}
\end{equation}
Repeating the same procedure leads to the same result in Eq.\,\eqref{eq:omeganumu_universal}, but without the factor $\pi (\mathcal{B})/ \pi(\bar{\mathcal{B}}) $. This finishes the proof that the mixed affinities can be expressed as Eq.\,\eqref{eq:coeffs} with $\omega_{\nu\mu}$ only depending on the kinds.

\begin{figure}
\centering
\includegraphics[scale=1]{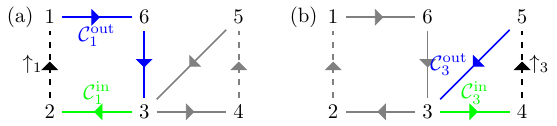}
\caption{Decomposition of cycles as per Eq.\,\eqref{eq:decompositionbridge} in the case where there is a bridge $\mathcal{B}$ connecting the two cycles (in this case it only contains state 3), by considering kinds 1 and 3. (a) Decomposition for the cycle $\mathcal{C}_1$ associated to the observable transition of kind 1 along $\up_1$. (b) Decomposition for the cycle $\mathcal{C}_3$ associated to the observable transition of kind 3 along $\up_3$.}
\label{fig:kinds13}
\end{figure}

\begin{figure}
\centering
\includegraphics[scale=1]{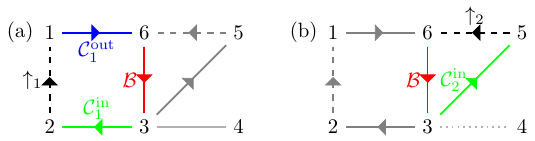}
\caption{When kinds 1 and 2 are considered, the cycles $\mathcal{C}_1$ and $\mathcal{C}_2$ overlaps over edge $\mathcal{B}$. The decomposition Eq.\,\eqref{eq:decompositionshared} has now to be used on the cycles. (a) Eq.\,\eqref{eq:decompositionshared} for cycle $\mathcal{C}_1$. (b) Eq.\,\eqref{eq:decompositionshared} for cycle $\mathcal{C}_2$. Notice that $\mathcal{C}_1^{\rm out}$ does not appear as the target state $\mathtt{t}(up_2)$ belongs to $\mathcal{B}$.}
\label{fig:kinds12}
\end{figure}

\section{}

\label{app:4}

Here we prove that the initial distribution $\boldsymbol{p}_1^* \propto D^{-1} \boldsymbol{1}$ satisfy
\begin{equation}
E(-\boldsymbol{a}) \boldsymbol{p}_1^* = J p_1^{*}.
\label{eq:prefrelation}
\end{equation}
We verify Eq.\,\eqref{eq:prefrelation} in the general case where a FR is obtained by tracking both total currents and mixed currents and when only total currents are tracked, that is the case for complete sets of currents.

For the first case we employ the solution Eq.\,\eqref{eq:preferredprobability}. Letting $Z_{\rm nc}$ denote the normalization of $p_{1,{\rm nc}}^*$
\begin{equation}
\fl \left[  E(-\boldsymbol{a}) \boldsymbol{p}_{1,{\rm nc}}^*\right]_{\ell_\nu} = \frac{1}{Z_{\rm nc}}\sum_{\ell_\mu'} e^{-a_\nu j(\ell_\nu)} \delta_{\ell_\nu,\ell_\mu'} e^{\frac{1}{2} a_\nu j(\ell_\nu)} = \frac{1}{Z_{\rm nc}} e^{\frac{1}{2} a_\nu j(\bar{\ell}_\nu)} =  p_{1,{\rm nc}}^* (\bar{\ell}_\nu) = \left[ J \boldsymbol{p}_{1,{\rm nc}}^* \right].
\end{equation}

In the second case, by using the solution Eq.\,\eqref{eq:preferredcomplete} for complete sets of currents, and letting $Z_{\rm c}$ denote the normalization of $p_{1,{\rm c}}^*$
\begin{equation}
\fl \left[  E(-\boldsymbol{a}) \boldsymbol{p}_{1,{\rm c}}^*\right]_{\ell_\nu} \frac{1}{Z_{\rm c}} \sum_{\ell_\mu'} e^{-a_\nu j(\ell_\nu) } \delta_{\ell_\nu,\ell_\mu'} e^{\frac{1}{2} a_\nu j(\ell_\nu) - U_\nu} \frac{1}{Z_{\rm c}} e^{\frac{1}{2} a_\nu j(\bar{\ell}_\nu) - U_\nu} = p_{1,{\rm c}}^* (\bar{\ell}_\nu) = \left[ J \boldsymbol{p}_{1,{\rm c}}^* \right].
\end{equation}

\section*{References}

\bibliographystyle{unsrt}

\bibliography{biblio}

\begin{thebibliography}{10}

\bibitem{beatofacurrent}
Pedro~E. Harunari, Alberto Garilli, and Matteo Polettini.
\newblock Beat of a current.
\newblock {\em Phys. Rev. E}, 107:L042105, Apr 2023.

\bibitem{Andrieux_2007}
David Andrieux and Pierre Gaspard.
\newblock Fluctuation theorem for currents and schnakenberg network theory.
\newblock {\em Journal of Statistical Physics}, 127(1):107--131, feb 2007.

\bibitem{schnakenberg}
J.~Schnakenberg.
\newblock Network theory of microscopic and macroscopic behavior of master equation systems.
\newblock {\em Rev. Mod. Phys.}, 48:571--585, Oct 1976.

\bibitem{avanzini2023methods}
Francesco Avanzini, Massimo Bilancioni, Vasco Cavina, Sara~Dal Cengio, Massimiliano Esposito, Gianmaria Falasco, Danilo Forastiere, Nahuel Freitas, Alberto Garilli, Pedro~E. Harunari, Vivien Lecomte, Alexandre Lazarescu, Shesha G.~Marehalli Srinivas, Charles Moslonka, Izaak Neri, Emanuele Penocchio, William~D. Piñeros, Matteo Polettini, Adarsh Raghu, Paul Raux, Ken Sekimoto, and Ariane Soret.
\newblock Methods and conversations in (post)modern thermodynamics, 2023.

\bibitem{transient}
Matteo Polettini and Massimiliano Esposito.
\newblock Transient fluctuation theorems for the currents and initial equilibrium ensembles.
\newblock {\em Journal of Statistical Mechanics: Theory and Experiment}, 2014(10):P10033, oct 2014.

\bibitem{Esposito_2012}
Massimiliano Esposito.
\newblock Stochastic thermodynamics under coarse graining.
\newblock {\em Physical Review E}, 85(4), apr 2012.

\bibitem{Bo_2017}
Stefano Bo and Antonio Celani.
\newblock Multiple-scale stochastic processes: Decimation, averaging and beyond.
\newblock {\em Physics Reports}, 670:1--59, feb 2017.

\bibitem{Andrieux}
David Andrieux and Pierre Gaspard.
\newblock A fluctuation theorem for currents and non-linear response coefficients.
\newblock {\em Journal of Statistical Mechanics: Theory and Experiment}, 2007(02):P02006--P02006, feb 2007.

\bibitem{hartich2014stochastic}
David Hartich, Andre~C Barato, and Udo Seifert.
\newblock Stochastic thermodynamics of bipartite systems: transfer entropy inequalities and a maxwell’s demon interpretation.
\newblock {\em Journal of Statistical Mechanics: Theory and Experiment}, 2014(2):P02016, 2014.

\bibitem{shiraishi2015fluctuation}
Naoto Shiraishi and Takahiro Sagawa.
\newblock Fluctuation theorem for partially masked nonequilibrium dynamics.
\newblock {\em Physical Review E}, 91(1):012130, 2015.

\bibitem{rosinberg2016continuous}
Martin~Luc Rosinberg and Jordan~M Horowitz.
\newblock Continuous information flow fluctuations.
\newblock {\em Europhysics Letters}, 116(1):10007, 2016.

\bibitem{crooks2019marginal}
Gavin~E Crooks and Susanne Still.
\newblock Marginal and conditional second laws of thermodynamics.
\newblock {\em Europhysics Letters}, 125(4):40005, 2019.

\bibitem{marginal}
Matteo Polettini and Massimiliano Esposito.
\newblock Effective thermodynamics for a marginal observer.
\newblock {\em Phys. Rev. Lett.}, 119:240601, Dec 2017.

\bibitem{effectivefluc}
Matteo Polettini and Massimiliano Esposito.
\newblock Effective fluctuation and response theory.
\newblock {\em Journal of Statistical Physics}, 176(1):94--168, apr 2019.

\bibitem{harunari}
Pedro~E. Harunari, Annwesha Dutta, Matteo Polettini, and \'Edgar Rold\'an.
\newblock What to learn from a few visible transitions' statistics?
\newblock {\em Phys. Rev. X}, 12:041026, Dec 2022.

\bibitem{vandermeer_2022}
Jann van~der Meer, Benjamin Ertel, and Udo Seifert.
\newblock Thermodynamic inference in partially accessible markov networks: A unifying perspective from transition-based waiting time distributions.
\newblock {\em Phys. Rev. X}, 12:031025, Aug 2022.

\bibitem{neri}
Izaak Neri and Matteo Polettini.
\newblock Extreme value statistics of edge currents in markov jump processes and their use for entropy production estimation.
\newblock {\em SciPost Physics}, 14(5):131, 2023.

\bibitem{kris-renewal}
Krzysztof Ptaszy\ifmmode~\acute{n}\else \'{n}\fi{}ski.
\newblock First-passage times in renewal and nonrenewal systems.
\newblock {\em Phys. Rev. E}, 97:012127, Jan 2018.

\bibitem{PhysRevLett.119.140604}
Simone Pigolotti, Izaak Neri, \'Edgar Rold\'an, and Frank J\"ulicher.
\newblock Generic properties of stochastic entropy production.
\newblock {\em Phys. Rev. Lett.}, 119:140604, Oct 2017.

\bibitem{Sokal1997}
A.~Sokal.
\newblock {\em Monte Carlo Methods in Statistical Mechanics: Foundations and New Algorithms}, pages 131--192.
\newblock Springer US, Boston, MA, 1997.

\bibitem{Budini}
Adrián~A Budini, Robert~M Turner, and Juan~P Garrahan.
\newblock Fluctuating observation time ensembles in the thermodynamics of trajectories.
\newblock {\em Journal of Statistical Mechanics: Theory and Experiment}, 2014(3):P03012, mar 2014.

\bibitem{VANDENBROECK2015}
C.~{Van den Broeck} and M.~Esposito.
\newblock Ensemble and trajectory thermodynamics: A brief introduction.
\newblock {\em Physica A: Statistical Mechanics and its Applications}, 418:6--16, 2015.
\newblock Proceedings of the 13th International Summer School on Fundamental Problems in Statistical Physics.

\bibitem{van2013stochastic}
Christian Van~den Broeck et~al.
\newblock Stochastic thermodynamics: A brief introduction.
\newblock {\em Phys. Complex Colloids}, 184:155--193, 2013.

\bibitem{douarche2005}
Fr{\'e}d{\'e}ric Douarche, Sergio Ciliberto, Artyom Petrosyan, and Ivan Rabbiosi.
\newblock An experimental test of the jarzynski equality in a mechanical experiment.
\newblock {\em Europhysics Letters}, 70(5):593, 2005.

\bibitem{Mossa2009}
A~Mossa, M~Manosas, N~Forns, J~M Huguet, and F~Ritort.
\newblock Dynamic force spectroscopy of {DNA} hairpins: I. force kinetics and free energy landscapes.
\newblock {\em Journal of Statistical Mechanics: Theory and Experiment}, 2009(02):P02060, feb 2009.

\bibitem{liphardt2002}
Jan Liphardt, Sophie Dumont, Steven~B. Smith, Ignacio Tinoco, and Carlos Bustamante.
\newblock Equilibrium information from nonequilibrium measurements in an experimental test of jarzynski's equality.
\newblock {\em Science}, 296(5574):1832--1835, 2002.

\bibitem{saira2012test}
O-P Saira, Y~Yoon, T~Tanttu, Mikko M{\"o}tt{\"o}nen, DV~Averin, and Jukka~P Pekola.
\newblock Test of the jarzynski and crooks fluctuation relations in an electronic system.
\newblock {\em Physical review letters}, 109(18):180601, 2012.

\bibitem{labbe1996power}
R~Labb{\'e}, J-F Pinton, and S~Fauve.
\newblock Power fluctuations in turbulent swirling flows.
\newblock {\em Journal de Physique II}, 6(7):1099--1110, 1996.

\bibitem{ciliberto2004experimental}
Sergio Ciliberto, Nicolas Garnier, S~Hernandez, C{\'e}drick Lacpatia, J-F Pinton, and G~Ruiz Chavarria.
\newblock Experimental test of the gallavotti--cohen fluctuation theorem in turbulent flows.
\newblock {\em Physica A: Statistical Mechanics and its Applications}, 340(1-3):240--250, 2004.

\bibitem{potts2019thermodynamic}
Patrick~P Potts and Peter Samuelsson.
\newblock Thermodynamic uncertainty relations including measurement and feedback.
\newblock {\em Physical Review E}, 100(5):052137, 2019.

\bibitem{horowitz2020thermodynamic}
Jordan~M Horowitz and Todd~R Gingrich.
\newblock Thermodynamic uncertainty relations constrain non-equilibrium fluctuations.
\newblock {\em Nature Physics}, 16(1):15--20, 2020.

\bibitem{suzuki2011}
Ryo Suzuki, Hong-Ren Jiang, and Masaki Sano.
\newblock Validity of fluctuation theorem on self-propelling particles.
\newblock {\em arXiv preprint arXiv:1104.5607}, 2011.

\bibitem{Gingrich_2016}
Todd~R. Gingrich, Jordan~M. Horowitz, Nikolay Perunov, and Jeremy~L. England.
\newblock Dissipation bounds all steady-state current fluctuations.
\newblock {\em Physical Review Letters}, 116(12), mar 2016.

\bibitem{Barato_2015}
Andre~C. Barato and Udo Seifert.
\newblock Thermodynamic uncertainty relation for biomolecular processes.
\newblock {\em Physical Review Letters}, 114(15), apr 2015.

\bibitem{nishiyama2002}
Masayoshi Nishiyama, Hideo Higuchi, and Toshio Yanagida.
\newblock Chemomechanical coupling of the forward and backward steps of single kinesin molecules.
\newblock {\em Nature Cell Biology}, 4(10):790--797, 2002.

\bibitem{evans1993}
Denis~J Evans, Ezechiel Godert~David Cohen, and Gary~P Morriss.
\newblock Probability of second law violations in shearing steady states.
\newblock {\em Physical review letters}, 71(15):2401, 1993.

\bibitem{evans1994}
Denis~J Evans and Debra~J Searles.
\newblock Equilibrium microstates which generate second law violating steady states.
\newblock {\em Physical Review E}, 50(2):1645, 1994.

\bibitem{touchette2009large}
Hugo Touchette.
\newblock The large deviation approach to statistical mechanics.
\newblock {\em Physics Reports}, 478(1-3):1--69, 2009.

\end{thebibliography}

\nocite{*}

\end{document}